%% file: chi_final_unmarked.tex
\documentclass[sigconf]{acmart}
\AtBeginDocument{%
  \providecommand\BibTeX{{%
    \normalfont B\kern-0.5em{\scshape i\kern-0.25em b}\kern-0.8em\TeX}}}
\usepackage{soul}
\usepackage{graphicx}
\usepackage{float} 
\usepackage{hyperref}
\usepackage{multirow}
\usepackage{tikz}

\usepackage{array}
\usepackage{cleveref}

\usepackage{xcolor}

\AtBeginDocument{%
  \providecommand\BibTeX{{%
    Bib\TeX}}}

\definecolor{babyblue}{rgb}{0.82, 0.96, 0.936} 

\newcommand{\DRone}[1]{
  \raisebox{-.5ex}{\hbox{\tikz{\node[fill=gray!20, rounded corners, inner sep=2pt] {\textcolor{black}{\textbf{#1}}};}}}
}

\newcommand{\DRtwo}[1]{%
  \raisebox{-0.5ex}{\hbox{\tikz{\node[fill=babyblue, rounded corners, inner sep=2pt] 
  {\textcolor{black}{\textbf{#1}}}
  ;}}}
}

\newcommand{\refone}[1]{%
  \raisebox{-.5ex}{\hbox{\tikz{\node[fill=gray!20, rounded corners, inner sep=2pt] {\textcolor{black}{\textbf{\hyperref[#1]{#1}}}};}}}
}%

\newcommand{\reftwo}[1]{%
  \raisebox{-.5ex}{\hbox{\tikz{\node[fill=babyblue, rounded corners, inner sep=2pt] {\textcolor{black}{\textbf{\hyperref[#1]{#1}}}};}}}
}%

\newcommand{\replace}[2]{\textcolor{black}{#2}}
\newcommand{\del}[1]{}
\newcommand{\final}[1]{{\textcolor{black}{#1}}}

\copyrightyear{2025}
\acmYear{2025}
\setcopyright{cc}
\setcctype{by}
\acmConference[CHI '25]{CHI Conference on Human Factors in Computing Systems}{April 26-May 1, 2025}{Yokohama, Japan}
\acmBooktitle{CHI Conference on Human Factors in Computing Systems (CHI '25), April 26-May 1, 2025, Yokohama, Japan}
\acmDOI{10.1145/3706598.3714398}
\acmISBN{979-8-4007-1394-1/25/04}

\begin{document}
\title{Gig2Gether: Data-sharing to Empower, Unify and Demystify Gig Work}

\author{Jane Hsieh}
\authornote{Both first authors contributed equally to this research.}
\affiliation{%
  \institution{Carnegie Mellon University}
  \city{Pittsburgh, PA}
  \country{USA}
}

\author{Angie Zhang}
\authornotemark[1]
\affiliation{%
  \institution{University of Texas at Austin}
  \city{Austin, TX}
  \country{USA}
  }

\author{Sajel Surati}
\authornote{Second authors also contributed equally to this work.}
\affiliation{%
  \institution{Bowdoin College}
  \country{USA}
  \city{Brunswick, Maine}
  }

\author{Sijia Xie}
\authornotemark[2]
\affiliation{%
  \institution{Carnegie Mellon University}
  \city{Pittsburgh, PA}
  \country{USA}
}

\author{Yeshua Ayala}
\affiliation{%
  \institution{Washington University in St. Louis}
  \city{St. Louis, MO}
  \country{USA}
}

\author{Nithila Sathiya}
\affiliation{
\institution{University of Texas at Austin}
\city{Austin, TX}
\country{USA}
}

\author{Tzu-Sheng Kuo}
\affiliation{
  \institution{Carnegie Mellon University}
  \city{Pittsburgh, PA}
  \country{USA}
}

\author{Min Kyung Lee}
\affiliation{%
  \institution{University of Texas at Austin}
  \city{Austin, TX}
  \country{USA}}

\author{Haiyi Zhu}
\affiliation{%
  \institution{Carnegie Mellon University}
  \city{Pittsburgh, PA}
  \country{USA}
}
\renewcommand{\shortauthors}{Hsieh \& Zhang et al.}

\begin{abstract}
The wide adoption of platformized work has generated remarkable advancements in the labor patterns and mobility of modern society. Underpinning such progress, gig workers are exposed to unprecedented challenges and accountabilities: lack of data transparency, social and physical isolation, as well as insufficient infrastructural safeguards. Gig2Gether presents a space designed for workers to engage in an initial experience of voluntarily contributing anecdotal and statistical data to affect policy and build solidarity across platforms by exchanging unifying and diverse experiences. Our 7-day field study with 16 active workers from three distinct platforms and work domains showed existing affordances of data-sharing: facilitating mutual support across platforms, as well as enabling financial reflection and planning. Additionally, workers envisioned future use cases of data-sharing for collectivism (e.g., collaborative examinations of algorithmic speculations) and informing policy (e.g., around safety and pay), which motivated (latent) worker desiderata of additional capabilities and data metrics. Based on these findings, we discuss remaining challenges to address and how data-sharing tools can complement existing structures to maximize worker empowerment and policy impact.
\end{abstract}



\begin{teaserfigure}
    \vspace{-5mm}
    \centering
    \includegraphics[width=.93\linewidth]{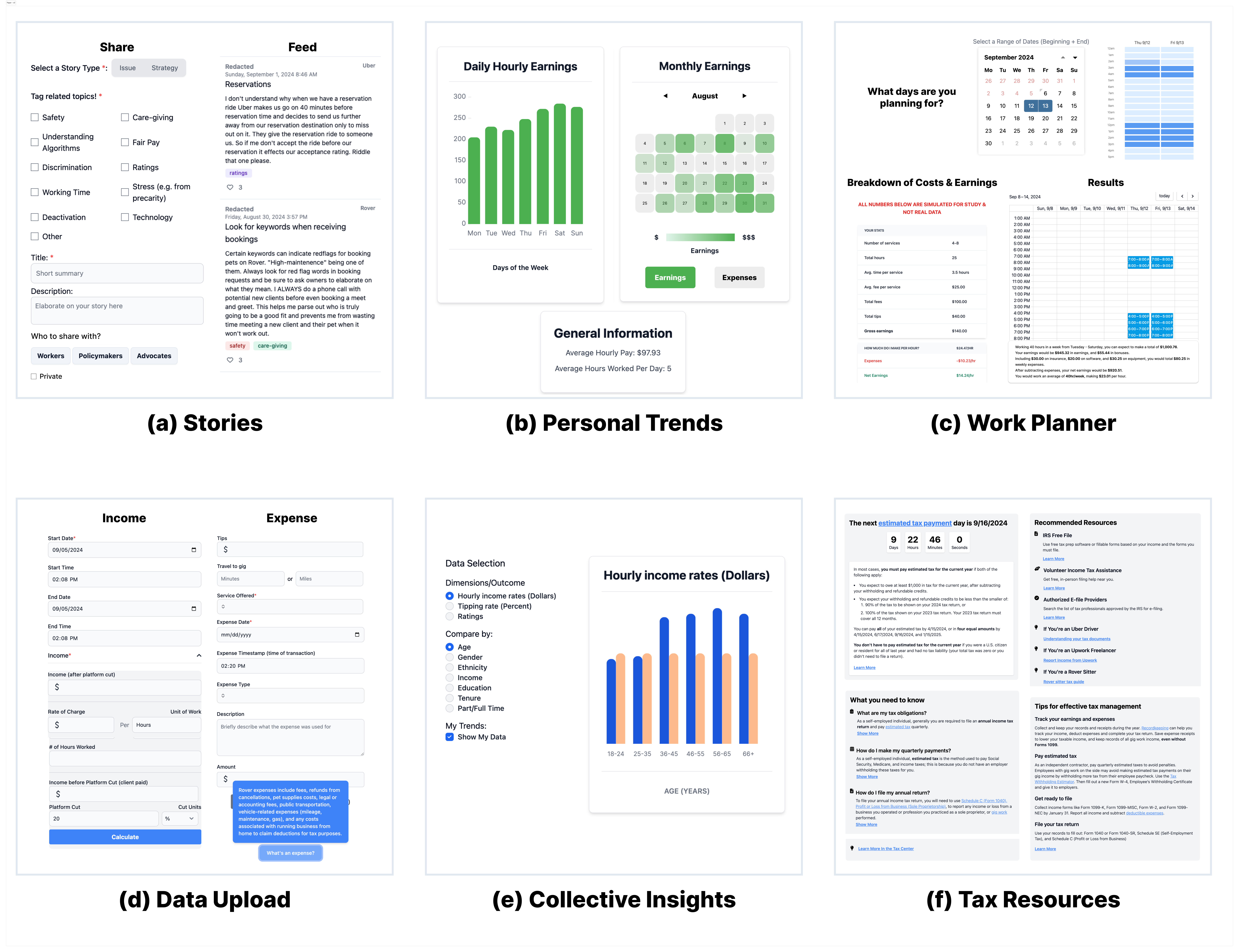}
    \vspace{-9mm}
    \caption{Gig2Gether features engage workers across platforms in qualitative data exchange via the Story Sharing and Feed (a), on top of financial tracking and datasharing via Income and Expense Uploads (d) Personal Trends (b), Collective Insights (e), Work Planner (c) and Tax Prep Resources (f)}
    \label{overview}
\end{teaserfigure}

\maketitle
\section{Introduction}
The rise of {platform-based gig} work over the past decade brought tremendous transformations {in} how society labors, hires, and commutes. Characterized by flexible, short-term contracts and project-based compensation \cite{looking}, gig work spans a variety of domains --- from transportation-based services like ridesharing and food delivery \cite{impact} to caretaking work such as petsitting and childcare \cite{beyond}, to fully remote tasks like freelancing and crowd work \cite{kuhn2019expanding}. 
By simplifying job matching and onboarding for workers while offering consumers a just-in-time workforce, {platformized gig work} bolsters the economy with ample supply of {convenient} services {and} a pool of frictionless job opportunities. 

However, underlying such transformational advances are major disruptions to how workers labor, and such experiences vary between gig platforms \cite{lived, privacy, lived_quality}. 
Workers of on-demand and physical work contexts (e.g., ridesharing, caretaking services) are regularly subject to subminimal pay \cite{subminimum, uberhappy}, stress, overwork, prolonged hours \cite{excessive, risks, stressfulride}, as well as fraud and harassment \cite{consent, brush} --- resulting in compromised worker safety and health \cite{health, safety}. Online freelancers and crowdworkers similarly experience financial precarity \cite{apouey2020gig, graham2017risks, toxtli2021quantifying}, emotional strain \cite{sousveillance, making}, and long and irregular hours \cite{wood2019good, lehdonvirta2018flexibility}. Care workers navigate job precarity \cite{caremodels, nurses}, wage theft \cite{ming2024wage, beyond}, and risks of entering strangers' homes \cite{bajwa2018health}.
On top of service variations, the gig picture is further complicated by the types of underlying algorithms \cite{looking}: on-demand services use automatic dispatch \cite{classification}, ranking-based systems allow clients to choose among sorted candidates \cite{context}, while bidding mechanisms oblige workers to apply to projects \cite{personal}. 
\del{This work engaged with workers from Uber, Rover, and Upwork to include the service domains of ridesharing, petcare, and online freelancing, \final{to study} the three variations of matching algorithms.} 

{Regardless of work contexts or underlying matching mechanisms, gig platforms accumulate power and control over workers through a similar strategy: extensive tracking of individuals' personal work data while intentionally withholding access to such data.} As both service providers and data producers, gig workers perform \textit{dual value production} for platforms \cite{dual}, who collect both service and data from workers to accumulate power and coerce labor, without assuming responsibility for direct and second-order consequences, risks and responsibilities \cite{taming}. 
Despite platforms' surveillance-style data collection and core reliance on workers' personal data to power algorithms, access to such data by non-platform stakeholders (e.g., advocates, policymakers, workers themselves) remains out of reach \cite{deficit}. Meanwhile, untamed platform algorithms widen and reinforce social inequalities \cite{ethical, profit}, in addition to keeping gig communities fragmented, divided and siloed \cite{atom, commodified, alienated} --- suppressing formation of a collective voice \cite{reinvent, rights, divided}.
{Such shortage of data access and a collective identity} critically impedes progress for labor regulation and worker classification in the gig sector, both of which lag behind in the US \cite{regulate, hamilton}. 

\final{Notably, prior work highlighted the propensity of policymakers toward generalized, all-encompassing policies for regulating gig work \cite{hsieh2023designing}. However, workers from understudied domains (e.g., caretaking \cite{undercare, care_advocacy}) often struggle with the normative tensions that affect contemporary labor policies \cite{rummery2012care, ferrant2014unpaid}. Thus, the inclusion of worker opinions from a diversity of domains is necessary to prevent further marginalization of laborers who offer often overlooked but critical services to society. 
On the other hand, separately studying policy priorities of workers from particular platforms can (1) preclude the potential of building collectivism and unification across platforms and (2) impose risks of duplicated efforts, as well as further disempowerment and segregation of worker voices that deviate from normative narratives.}
To empower and elevate both diverse and unifying gig work experiences, this study proposes a system catering to workers from varying platforms to support cross-domain worker interactions.

The need for {sharing information and building solidarity} among gig workers prompted calls from the HCI {and CSCW communities} to build worker-centered data collectives \cite{supporting, cscw_workshop, calacci2022organizing, stein2023you, sousveillance}. 
But despite the shared underlying challenges that hinder gig worker collectivism and the recognition among the research community of the necessity of collective data-sharing to inform evidence-driven policy initiatives, the gig workforce remains divided and segregated \cite{supporting}.
Further, most gig workers lack practical experience in intentionally contributing personal data for purposes of building collectivism or informing policy initiatives, leaving an open question of whether workers themselves would be motivated to engage in data-sharing. Despite existence of prior systems that focused on building collectivism through data/experience-sharing within specific gig work communities \cite{dynamo, calacci2022bargaining}, we are unaware of research exploring how workers engage with a cross-platform data-sharing tool situated within their everyday workflows.

{We took a first step to address this gap by building and evaluating a prototype data-sharing system aimed to connect active workers from three gig platforms.
Based on design requirements {(\S\ref{Related_Work_Design})} derived from related literature, we constructed early wireframes {(\S\ref{early})} and
conducted pilot testing with workers in our target domains (\S\ref{iterative}) to ensure alignment with worker preferences and refine usability.}
{This multi-stage design process culminated in} Gig2Gether (\S\ref{gig2gether}), a {prototype} system {enabling} US-based workers {from three gig platforms} to (1) {engage in cross-platform mutual support through data- and experience-sharing to promote larger-scale solidarity and cooperation}, (2) track {and reflect on work experiences and statistics \final{regarding personal and aggregate} working conditions}, and (3) {strategize and} plan {for improving their gig careers}. {Beyond} {surfacing possible cross-platform} worker {interactions, reactions and unfulfilled }desiderata, Gig2Gether is intended to eventually serve as a portal {for exchanging knowledge, insights and resources} between workers, advocates and policymakers.

Through a subsequent field study {(\S\ref{field})} with 16 gig workers across three platforms/work domains, we {surfaced three main themes around how gig workflows can integrate data-sharing for empowering collectivism and advancing policy.} 
{First, the exchange of experiential strategies and challenges allowed workers to engage in cross-platform mutual aid, 
individual tracking of financial data enabled them to reflect on and plan work,
while potential shared tracking of aggregated statistics helped them imagine use cases of collaboratively reasoning about platform mechanisms and rates. Second, workers expressed willingness to share both aggregated statistics and qualitative accounts of lived experiences with other stakeholders, for purposes of \final{motivating} policy creation, especially regarding safety and wages. Third, we overview how data-sharing integrated into workers' varied workflows, describing practical challenges that inform desires of future affordances as well as requests for additional metrics and data.} Finally, we discuss and reflect on {additional practical challenges unveiled (\S\ref{ongoing})}, potential implication for {advocacy and policy influence (\S\ref{door}), as well as ways that data-sharing can complement existing and alternative means of worker empowerment (\S\ref{complement})}.

\section{Related Work}\label{Related_Work}
{Recent efforts} surrounding the investigation and improvement of gig work conditions coalesced around 1) the potential of worker data for making evident the conditions imparted by algorithmic platform practices and 2) the importance of concrete policy and regulation that ensure strong worker protections. 
Below, we describe related works that center our vision {and design} of a worker-centered data-sharing platform that meets {the needs of workers (across platforms)} for self-tracking, mutual aid, and policy advancements that improve current work conditions.

\subsection{{Demands of Gig Work: Shared and Divergent Challenges Across Platforms}}
\input{chi-related-work-2.1.tex}

\subsection{{Work Tracking: Individual Logging $\rightarrow$ Collective Support \& Decision-making}}
In the absence of peer support and higher power actors who assume or share the structural risks and challenges inherent to gigs, workers are left to their own devices to manage {various} accountabilities {and obligations} \cite{consent, indie, accountable}. 
Studies documented two main ways that workers understand and manage work: on their own through self-tracking, or with peers via online groups/forums. 
Recent work at the intersection of HCI and Personal Informatics revealed how gig workers currently (or might in the future) self-track to (1) protect themselves from the platform \cite{privacy} or customers \cite{visibility, sousveillance} (2) comply with tax obligations \cite{tax_lives, taxing} (3) understand how algorithms operate \cite{sousveillance, zhang2023stakeholder} and (4) comprehend and improve their own earning patterns \cite{zhang2023stakeholder, accountable, supporting} using tools such as data probes {in addition to apps designed for tracking fuel, time, tax, mileage \footnote{Tracking apps include Fuelio and GasBuddy (for Fuel), Traqq (for time), Stride (for tax), as well as MileIQ, Everlance and Triplog (for Mileage)} and generalized gig work assistance \cite{accountable}. For instance, Mystro a commercial tool affording rideshare drivers the agency to auto-decline work across platforms that do not match their expressed preferences (e.g., earning rates, duration of gigs, work locations). 
Gridwise and Farepilot provide workers data-driven insights about in-demand locations, while Stride assists with tax filing. } 

While such tools act as resource providers and \del{(sometimes automated) }assistants, they {fall short in} provid{ing} workers with social support or strategies in times of need. T\del{hus, t}o overcome \replace{social isolation 
}{the atomized nature gigs \cite{peersupport, atom}} and find a sense of   ``community'',  workers also {}leverage online forums (both {pages and groups on general-purpose sites like Reddit/Facebook} and platform-specific sites like uberpeople.net) to share strategies \cite{nomad} and information \cite{machines, peersupport} {so they can hypothesize and collective make sense of underlying platforms' algorithmic mechanisms}, solicit advice and social connections \cite{belonging, organizing}, as well as rant and commiserate \cite{privacy, atom}.  
Online video tutorials (e.g., vlogging) are also emerging as more effortless ways for workers to learn about existing strategies and work conditions \cite{chan2019becoming, pires2024delivery, woodside2021bottom}.
However, the loosely-organized structure of general purpose forums (and video sharing platforms) {makes them in}effective for sensemaking \cite{peersupport}, while platform-specific sites limit {worker's abilities to discern unifying challenges shared across domains from characteristics that uniquely afflict workers of a single work context/platform}.
{Furthermore, ``Online forums are built to aid workers with a sense of immediacy, not to quantifiably or qualitatively monitor
request patterns or worker grievances over time'' \cite{organizing}, making them ill-suited for purposes of collective bargaining or identify-building.}
\vspace{-1mm}
\input{chi_2.3}
\vspace{-1mm}
\section{Research Team \& Positionality}
Responding to calls urging researchers to go beyond stating our findings, observations and analysis as passive and objective knowledge \cite{kirsch1999ethical, turkopticon}, we state our own backgrounds and trainings as researchers to reflect on our own identities, assumptions and values --- so as to make sure we acknowledge, reflect on and avoid risks of unintentionally substituting our own voice over those of our participants \cite{koelle2020social, borning2012next}, as well as to recognize our own relatively-advantaged positions and experiences as designers and researchers -- privileges that our participants do not have. 
Our team members specifically work and receive training in the fields of Human-Computer Interaction, Software Engineering, Information Studies, Integrated Innovation, Computer Science, and Mathematics at four universities located in the US. 
Four authors have prior research experience engaging with gig workers across platforms, and another author currently works part-time as a gig worker on Rover, but we continue to acknowledge the power imbalance induced by the research-participant relationship. 
To address this disparity, as well as maintain a transparent and worker-centric procedure, we (1) revealed to worker participants our intentions to aid workers and inform policy, and (2) made sure they understood the third-party nature of our investigations (independent of gig platforms).
\vspace{-1mm}
\section{Design {Space \&} Requirements for Worker-Centered Data-sharing Tools}
\label{Related_Work_Design}

We first outline the worker populations that we aspire to impact with a data-sharing collective to delineate the scope of our design space and context(s) of study. Then, we draw insights and findings from prior works around gig worker-led data-sharing to formulate two design requirements that guided the development of our tool.

\paragraph{{Target User/Worker Populations}}
{To support gig workers in building collective identity and offering mutual aid to peers experiencing similar values and priorities, we intend to impact gig workers from a variety of platforms in our design of a worker-centered data-sharing tool. In particular, we consider individuals to be a part of this intended audience if they primarily use algorithmically-managed gig platforms to find and conduct both physical and remote work. }
 
While we recognize the populations of workers who complete gigs but do not use gig platforms to procure them --- e.g., contractors belong to LLC's or other small businesses, as well as artists or musicians who leverage other means of networking to acquire gigs --- we do not consider such groups to be under the scope of this study, since their job acquisition process do not require individual workers to interact with a gig platform as an algorithmic intermediary.

\final{Finally, while we broadly aspire to unite workers of varying domains, we also recognize potential differences in background and experiences that may create disparities in how they perceive or use a data-sharing tool. In this initial evaluation, we do not focus on individual worker experiences that go beyond variations of work domains, but urge ongoing and future work to consider differences.}

\subsection{{Design Requirement 1}: Center Worker {Needs \&} Goals {for Advancing Policy}} \label{DR1}
Related studies exploring designs of collective worker data-sharing tools approached the issue with worker-centered and participatory design methods \cite{stein2023you, zhang2023stakeholder, supporting}. Several suggested that identifying and accommodating worker needs requires both statistical and contextual data around work conditions \cite{hsieh2023designing, zhang2024data}, shared and collected in a way that meets current worker goals and workflows, while ensuring agency (i.e. data control) over what they share, how often they share, and who they share to {\cite{supporting, stein2023you}}. In the following sub-requirements, we detail some surfaced needs to respect existing worker habits and preferences while supporting data contributions that promote self-assessment and policy advancements.


\subsubsection*{\DRone{DR 1.1}: \textbf{Support Quantitative and \underline{Qualitative} Data Sharing} {for Impacting Policy}}
\label{DR1.1}
As \S\ref{Related_Work_Challenges} details, 
gig working conditions are riddled with issues that vary across platforms, although ``even workers on the same platform can experience \dots differences'' \cite{supporting}. 
While some challenges (e.g., low and unfair pay \cite{calacci2022bargaining}, long and irregular hours \cite{lehdonvirta2018flexibility}) can be observed through quantitative data, many other factors that critically contribute to unpaid/invisible labor --- e.g., emotional stressors in care work \cite{ming2023go, supporting}, discrimination \cite{disability}, compromised safety standards \cite{taylor2023physical} --- require qualitative forms of data to descriptively report the issue within its applicable contexts
\del{While quantitative data help stakeholders directly measure effects of algorithmic management on outcomes such as hours of engagement and pay \cite{calacci2022bargaining}, 
narrative accounts help generate \cite{supporting}, document and raise awareness around new, nuanced and contextual factors that contribute to invisible labor and hidden risks} \cite{lived}, 
especially given the rapidly-evolving nature of platform policies and algorithms \cite{pacifying}. 
In particular, \citet{zhang2024data} highlighted the potential of worker-centered tools to ``spotlight workers’ lived experiences'' and bring oversights in labor regulation ``to the attention of regulatory bodies''. Thus, an effective data-sharing tool should provide avenues for quantitative and qualitative contributions. 

\subsubsection*{\DRone{DR 1.2}: \textbf{Build Trust via Privacy, Security, and Data Control}}
\label{DR1.2}
Within online communities where identity disclosures are optional, establishing trust is prevalent and ongoing problem \cite{cmc, dynamo}. To ensure safety and security for workers who contribute personal data, a data-sharing tool should be equipped with appropriate security precautions and policies, as well as configurable privacy options. 

In prior investigations, gig workers prioritizing \textbf{security} concerns cited fears of ``backlash, harming reputation, and legal consequence[s]'' \cite{sousveillance} from platforms for actions like ''breaking platform terms of service'' or retaliation tactics such as ``shadow bans'' \cite{stein2023you}, while others worried about releasing locational data to ``past problematic clients'' \cite{supporting}. To minimize  security risks, some recommended techniques like ``anonymization, aggregation and perturbation of data'' \cite{stein2023you}, atop ways for workers ``to revoke data access'' \cite{supporting}. Workers also generally prefer sharing ``aggregate data but not individual data with peers, primarily due to concerns related to competition'' \cite{supporting}.
To this end, a data-sharing tool should anonymize all collected forms of quantitative data, in addition to providing anonymity-preserving sharing options for qualitative data. No worker accounts should have permission to view identifiable personal work data of other workers. 
Furthermore, workers should have control over the granularity of detail for each piece of uploaded data (how much), length of data persistence (for how long), who they share their data with (to whom), as well as whether they contribute quantitative or qualitative data (what to share). 

Around \textbf{privacy}, workers found ``trust [to] go hand in hand with privacy policies'' \cite{supporting}, therefore a data-sharing tool should remain transparent about how uploaded data get used by the system.
We note that despite the close ties of privacy to trust, attempts at eliciting privacy requirements uncovered a paradox where although ``workers were aware of the risks of sharing data'' \cite{stein2023you, privacy}, they ``were largely unconcerned with their likelihood'' \cite{supporting, stein2023you}, suggesting that without a working prototype of a data-sharing system simulating the in-situ experience of contributing and uploading on a daily basis, ''it can be difficult to imagine and consider related concerns with data privacy'' when workers lack ``practical experience engaging with civic tech or data activism''. This further underscores the importance of transparently disclosing to workers the types of data collected and how it gets used by the system.
\vspace{-1mm}
\subsubsection*{\DRone{DR 1.3}: \textbf{Support Heterogeneous Goals \& Workflows}}
\label{DR1.3}
Prior investigations found variances in workflows and goals among workers, creating ``divergent preferences on how to best upload data'' \cite{supporting} and ``no consistency on the types of data'' to upload \cite{stein2023you}. 
\citet{accountable} additionally found workers to integrate ``multiple tracking tools''
for income tracking and planning in their work routine to ``learn what the job is like, determine if their jobs are worth continuing, know how much they’re earning, monitor productivity, and manage work/life balance''.
Although workers should not be required or encouraged to use its every available feature, the system should provide workers with multiple data upload methods, a variety of worker-centered features to support different incentives, as well as incorporate and centralize financial tracking features.
For example, while some Uber drivers might be curious about their estimated earnings for particular Quests, others might simply want to track their earnings per trip\cite{zhang2023stakeholder} --- workers should have methods for keeping track of both units of work. 
\vspace{-1mm}
\subsection{DR 2: Facilitate Worker Collaboration \& {Cross-Stakeholder} Resource Sharing}
\label{DR2}
While gig workers engage with existing online forums \cite{atom} and self-tracking tools \cite{accountable} -- where they exchange experiential knowledge to support learning/understandings of platforms and work conditions -- we are unaware of online space(s) designed specifically for workers across gig platforms and domains to contribute to a shared data repository, or that connect workers across platforms to existing resources. Below, we present four guidelines for creating digital environments that foster worker collectivism and organize resources of benefit and use to the general gig workforce.

\subsubsection*{\DRtwo{DR 2.1}: \textbf{Encourage Contributions to Inform Labor Initiatives}}
\label{DR2.1}
While prior works \cite{supporting, hsieh2023co, zhang2023stakeholder} identified {shared}
 concerns around gig work that both policy experts and workers considered priorities (e.g. equity, fair pay, safety) data surrounding those topics are scarce to nonexistent, due to platforms' reluctance to share. 
To rectify this data deficit, \citet{supporting} recommended using qualitative data such as ``personal anecdotes'' to pinpoint drivers of discrimination, ``digestible breakdowns of costs and earnings'' to educate and bring awareness to workers (and the public at large) about whether they making above minimum wage, and ``communication channel between workers and policy experts'' to facilitate worker reports of power imbalances with clients via data such as ``cancellations and safety reports'' \cite{stein2023you}.

\subsubsection*{\DRtwo{DR 2.2}: \textbf{Connect Workers to Other Stakeholders' Resources}}
\label{DR2.2} 
{As self-employed individuals, gig workers shoulder several resource accountabilities (e.g., financial, network), in the absence of organizational support \cite{accountable}.}
In discussions with policymakers and advocates, \citet{hsieh2023co} received many pointers from organizations and advocates for resources targeted to gig workers{, including ``employee assistance and job training programs''}. Unfortunately, there is currently no centralized space for disseminating such information. 
Possibly driven by a fear of factors such as competition and spam content, gig workers are disincentivized from constructing open, Wikipedia-like portals where they collectively gather and use ``data, insights and contextualize information'' around work conditions \cite{stein2023you}.
A data-sharing tool can serve as a portal for connecting workers to such known resources.

\subsubsection*{\DRtwo{DR 2.3}: \textbf{Multi-Domain Support {\& Worker-Accessible Tools}}} 
\label{DR2.3}
{As described in \S\ref{Related_Work_Challenges}, gig work} span a variety of work domains \cite{impact,beyond,kuhn2019expanding}, making it crucial for a data-sharing tool reach workers of different services{, especially since ``different types of tasks gig workers engaged in affected their preferences on what gets shared, how it is submitted, and how often it is to be uploaded'' \cite{supporting}}. 
To accommodate heterogeneous workflows, needs and data types involved with varying gig domains, data-sharing systems should afford workers agency to customize sharing preferences --- e.g., formats of data to upload, and what devices to upload from. 
For instance, \citet{calacci2022bargaining} pointed out how workers performing physical services (e.g., grocery shopping) often ``do not own a desktop computer, so any solution had to be easily accessible from a mobile device'', but workers offering digital services (e.g., online freelancers) may prefer desktop solutions that embed into their existing workflows. Thus, data-sharing tools should respect device preferences of online and offline service providers\del{ should be accessible via both phones and laptops}.

\subsubsection*{\DRtwo{DR 2.4}: \textbf{Cross-Worker/Stakeholder Communication}}
\label{DR2.4}
{While the value and necessity of achieving ``effective representation and collective bargaining for workers in the gig economy'' is widely recognized \cite{calacci2023access, cscw_workshop, reinvent, rights}, }the online and individual nature of work isolates workers from peers \cite{atom, commodified, alienated}, making gig work collectivism the `holy grail' of the community. 
To truly connect workers in a network that benefits themselves instead of platforms \cite{commodified}, the tool should allow communication between gig workers, including across platforms.
Additionally, the system should also open up collaboration to higher-power stakeholders such as policymakers and advocacy groups {to ``find ways of maximizing their ability to support gig workers'' \cite{hsieh2023co}}. 
\begin{figure*}[h!]
    \centering
    \includegraphics[width=\textwidth]{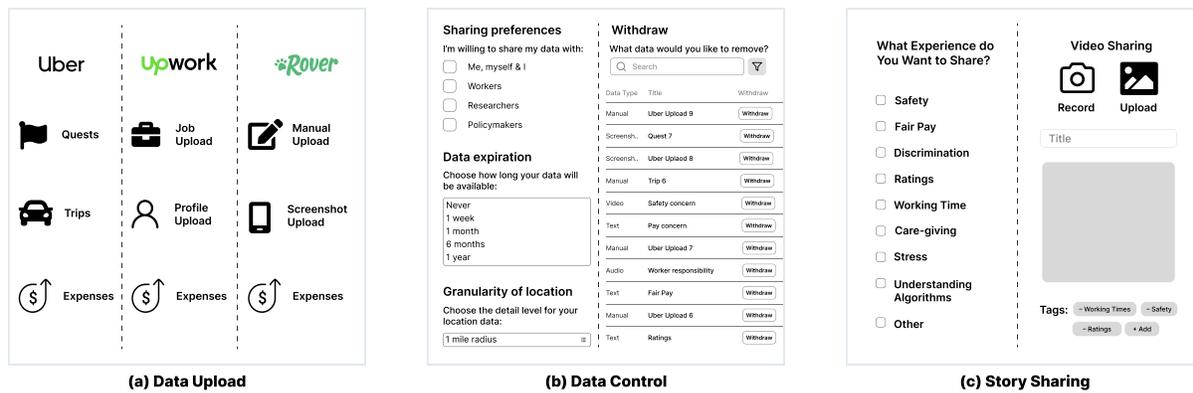}
    {\small
    \caption{Initial Sketches -- (a) aligns to \protect\refone{DR1.3} \& \protect\reftwo{DR2.3}, (b) accommodates \protect\refone{DR1.2} \& \protect\reftwo{DR2.2}, (c) targets \protect\reftwo{DR2.1} \& \protect\reftwo{DR2.3}}
    \label{sketches}
    }
\end{figure*}
\section{{Design Exploration of a Worker-Centered Datasharing Tool}}

The above design requirements guided our conceptualization of a first low-fidelity wireframe (see Figure \ref{sketches}) for a prototype system that shares worker data across stakeholder groups and gig platforms. 
For this study, we chose platforms that represent domains of petcare, rideshare driving, and online freelancing in an attempt to capture experiences from a diversity of work domains (see \S\ref{Related_Work_Challenges}).
Through think-aloud pilot tests with six workers, we solicited for and iteratively incorporated feedback {into} the wireframe{s} to further align our designs with worker objectives. 

\subsection{{Early Prototyping: Low-Fidelity Sketches}}\label{early}

Prior to implementation, we conducted initial brainstorming sessions and created low-fidelity sketches to conceptualize the tool's key features, guided by above design requirements\del{objectives derived from prior research and worker needs}.  To accommodate diverse worker goals and workflows \refone{DR1.3}, and varying work domains \reftwo{DR2.3}, we designed multiple data upload methods (see (a) of Figure \ref{sketches}).
\del{(a) to present Personal Trends, allowing workers to evaluate their performance and utilize the Planner feature for future work scheduling based on historical data predictions}
To meet \reftwo{DR2.4}, we envisioned the Story feature (c) and Collective Trends, which facilitate collective advocacy and mutual aid through both quantitative and qualitative data sharing \refone{DR1.1}, aided by granularized data controls that manage data visibility toward multi-stakeholder groups \reftwo{DR2.2}, and provide workers autonomy to withdraw data or hide identifiable information such as location (b) \del{, aligning with} \refone{DR1.2}.
The Story feature includes \del{ video-sharing capabilities and}\reftwo{DR2.1}-aligned tags to directly inform policy initiatives. We iteratively refined these initial concepts through sketching sessions and discourse within the research team to ensure the alignment of features with identified design requirements.

\subsection{Iterative Pilot Testing with Workers}
\label{iterative}

Leveraging think-aloud responses from initial sessions with three workers, we iteratively refined these wireframes into medium-fidelity prototypes. 
Worker feedback significantly shaped our design choices, particularly regarding data sharing and privacy. For instance, a petsitter expressed concerns about tag usage for identification, ``because people may easily find her'' but was ``willing to post stories with her initials.'' 
Workers also emphasized the importance of understanding the purpose of data sharing, with a driver requesting clarity on ``What are you looking for from us, What is the data going to do?'' 
Additionally, we incorporated a requested Tax Planner feature after several workers expressed interest in ``download[ing] information for taxes purposes.'' 
The final wireframes guided our implementation of the responsive web app for Gig2Gether, which included careful terminology adjustments (e.g., changing ``worker'' to ``gig worker'' based on feedback from pilots\del{that ``calling it 'workers' will confuse people''}). 
After developing a preliminary version of the functioning web app, we ran a second round of think-alouds with three other workers.

For both rounds of testing, we recruited three workers (one per domain\del{ between rideshare drivers, petsitters and freelancers}) --- resulting in six pilot testers in total, who were recruited as contacts from previous studies. Workers appreciated several features: one ``liked being able to track the expenses in the expense log; another highlighted that they ``like[d] understanding algorithms version of stories.'' 
Workers also suggested valuable feedback for improvements, such as implementing mid-day notification reminders for data uploads. The semi-structured think-aloud questions and tested mock-up examples are included in Appendix \ref{A.1.1} and \ref{thinkaloud}
\begin{figure*}
    \centering
    \includegraphics[width=\linewidth]{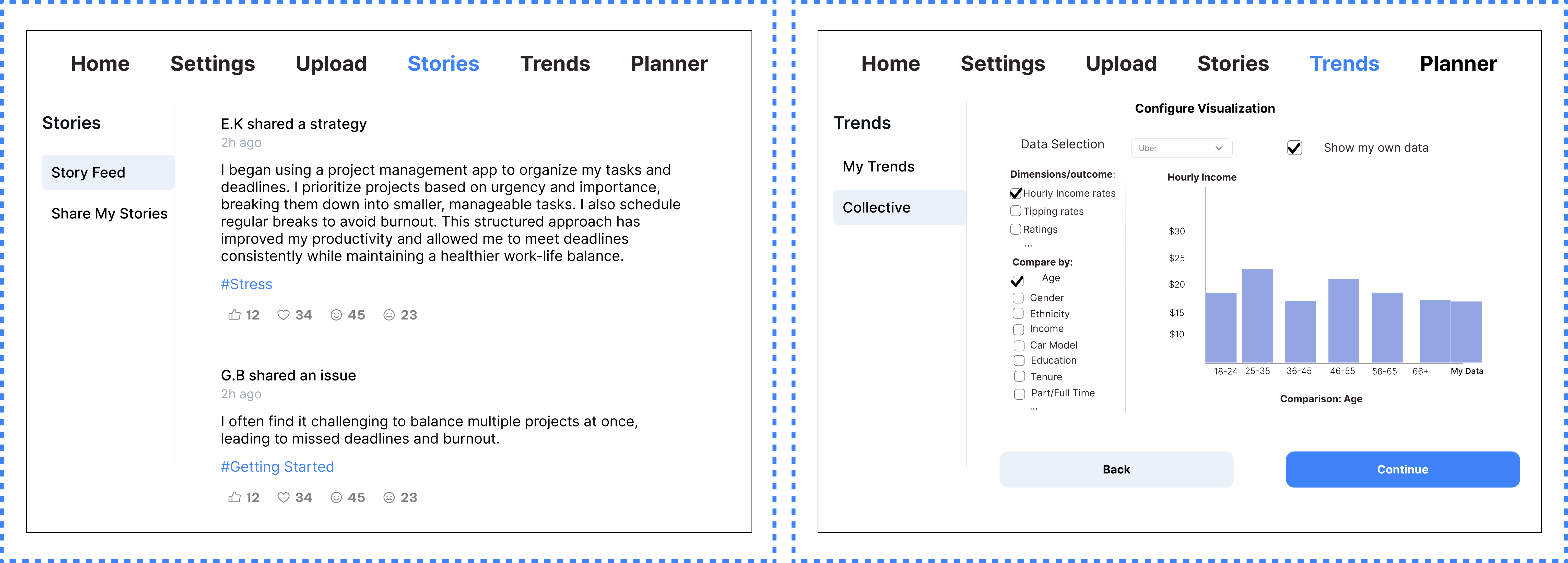}
    \caption{Web Versions of Mid-fidelity Wireframes}
    \label{mid_fidelity}
\end{figure*}

\section{Gig2Gether}\label{gig2gether}

Based on the mid-fidelity prototype (see Figure \ref{mid_fidelity}) generated from the iterative design process, we developed Gig2Gether: a worker-centered data-sharing tool with capabilities for uploading work data, viewing personal and collective work trends, sharing stories about work, as well as planning work and taxes. 
Built as a web app, Gig2Gether accommodates workers operating from various devices (laptop, mobile, \& any device with web-browsing capabilities). 
The app consists of a frontend built with SvelteKit and backend (database, storage and analytics) supported by Firebase.

\begin{figure*}[h]
    \centering
    \includegraphics[width=.8\textwidth]{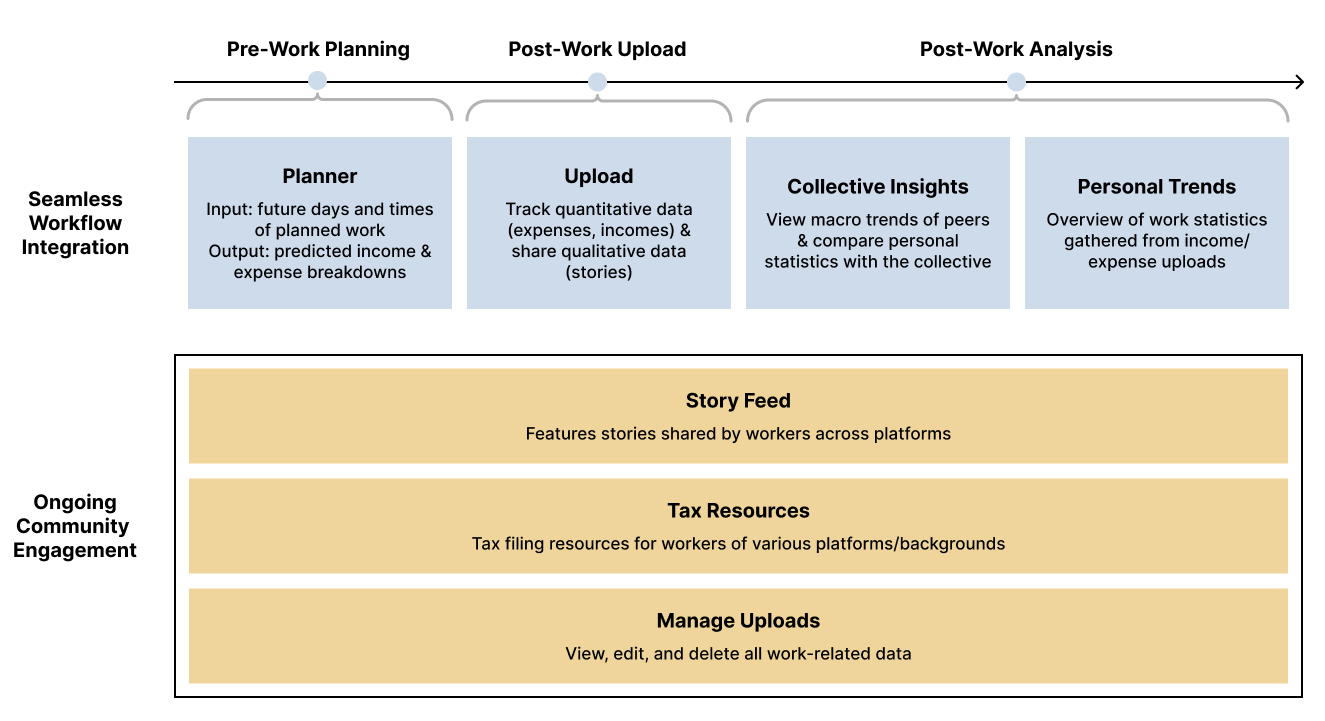}
    \caption{Overview of Gig2Gether Features for Before, After and Between Gigs}
    \label{timeline}
\end{figure*}

Users can leverage the system to plan for, record and reflect on work at various stages of a job.
Before a gig, workers can use the planner to predict future earnings and set work goals. After finishing a task, {workers can store and share its} associated earnings, expenses and stories.
After uploading data {for} the recently completed task, workers can view and reflect on personal work trends, or use collective insights to grasp macro-level statistics about comparable or contrasting worker populations. 
Between gigs, workers can leverage  the 1) \textit{story feed} to learn about strategies or recent work conditions reported by peers, 2) \textit{tax page} to peruse resources that support fulfillment of tax obligations or 3) \textit{profile page} to reflect on their history with the platform or record repeatedly incurring expenses. 
Figure \ref{timeline} illustrates this timeline while subsequent subsections briefly discuss each feature and associated design requirements. \final{More details of each feature are expanded on in Appendix \ref{details}}.

\subsection{Exchanging Qualitative Stories}
One of the intentions of Gig2Gether is to maintain a community for gig workers to share their own experiences with peers as well as policymakers and advocates, so as to help alleviate social isolation. To fulfill this objective, the Stories panel allows users to read and post about strategies and issues related to their everyday work. When sharing stories, workers are required to choose (1) related tag(s) --- prepopulated with themes identified from Section \ref{Related_Work_Design} --- (2) whether their Story represents an issue or strategy, and (3) who to share the Story to.

The \textit{Story} feature --- Figure \ref{overview}(a) --- serves to meet \reftwo{DR2.4} by enabling users to exchange data and strategies with peers, support others' stories, as well as share strategies and insights gathered surrounding platform policies and feature updates. 
To align with \refone{DR1.2}, workers have agency to configure desired viewing audiences of each shared post, and each Story is associated to authors with only via usernames. Tags encourage the sharing stories related to initiatives of interest to policymakers, in observance of \reftwo{DR2.1}.

\subsection{Upload Earning \& Expense Data}
\label{upload}
One key feature of Gig2Gether is to help workers keep track of data surrounding their gigs so they can remain financially accountable. \final{To support such financial tracking, workers can upload leverage both \textit{Income} and \textit{Expense} forms to monitor earnings --- Figure \ref{overview}(d).}
\del{Below, we outline how workers of the three domains/platforms can upload income and expense entries. }
In response to \refone{DR1.2}, income and expense uploads require only a small set of information: date, length and type of work, as well as income amount for income entries while expense entries only require data and amount of expense.
This way, workers retain agency over to choose the fields to share or track about income and expense entries. To further address the data control requirement, Gig2Gether provides manage data pages for users to view, modify and delete and story, expense, and income uploads at any point. 
In the income uploads, drivers have options to submit data manually or streamline the process by uploading their CSV's, in adherence to \refone{DR1.3}. Finally, the custom form fields of expense and income entries for each platform complies to multi-domain support (\reftwo{DR2.3}).

\subsection{Viewing Work Trends}
To provide workers more information about their own work, as well as insights on their standing among other gig workers, we created pages --- Figure \ref{overview}(e) --- for workers to view their own earnings data over time (\textit{Personal Trends}) as well as one to compare and contrast how their statistics fit within more aggregated-level data (\textit{Collective Insights}).
Both personal and collective trends map directly to the \refone{DR1.1}, and all inputs to eventually populate collective trends will be anonymized to protect workers privacy (\refone{DR1.2}).

\subsection{Planner}
Currently, Gig2Gether offers a prototype of a work planning feature, intended to inform its about predicted future earnings based on user inputs of planned work hours and historical data. The Planner --- Figure \ref{overview}(c) --- takes in a range of the days and hours a user plans to work, and displays a simulated summary and breakdown of what predicted earnings might look like. 
In the future, the Planner would populate the earning projections using users' historical data and work trends or patterns. 
Implementation of the Planner was guided by \refone{DR1.1} to help workers gain personal statistics, since predicted data is directly based on the user's history of uploaded information. The Planner drew from the design, inputs, and outputs of the Planner data probe in \cite{zhang2023stakeholder}.
\subsection{Additional Features}

\paragraph{Tax Preparation} In adherence of \reftwo{DR2.2},
the tax page --- Figure \ref{overview}(f) --- integrates resources for part/full-time workers, guides from platforms, as well as general information about filing. It tracks the next tax day for eligible workers, in addition to providing information, resources, and tax preparation tips.
\paragraph{Multi-Domain Support}
Story sharing allows for cross-domain communication between users, following \reftwo{DR2.4} to inspire a gig worker collective. Additionally, workers have access to a variety of tax resources for all gig work domains. This helps workers who work multiple types of jobs to reference domain-specific tax resources for all three gig platforms (supported by Gig2Gether).

\section{Field Evaluation Methods}\label{field}
To assess the practical application of Gig2Gether in the daily working lives of various gig workers and how it can assist them in gathering evidence of issues to share with policymakers, we conducted a field study with 14 gig workers across the three domains. Workers were asked to use the system for 7 consecutive days, in addition to 1-hour onboarding and exit interviews.
\input{participants}
\subsection{Recruitment}
We recruited gig workers through various channels, including r/Upwork, r/RoverPetSitting, and r/Uber subreddits. In addition to Reddit, we posted on city-specific Nextdoor and Craigslist, reached out to participants from prior studies and handed out flyers to Uber drivers in-person at airports. 
Interested individuals completed pre-screening surveys to ensure eligibility and diversity in work types, locations, and experience levels. 
Selected participants then completed a consent form and a pre-study questionnaire to gather demographic information. 
In total, we recruited 16 gig workers active on three platforms (8 Uber drivers, 5 Rover petsitters, and 2 Upwork freelancers) with varied experience levels, as shown in Table \ref{participants}. Onboarding sessions and exit interviews were conducted via Zoom. Participants received up to \$200 USD as compensation, including \$30 for onboarding, \$140 for the field study (\$15 per day for 7 days plus \$35 for optional tasks) and \$30 for the exit interview.
\subsection{Onboarding Interviews}
The field study commenced with a one-hour, one-on-one onboarding session. At the beginning of each onboarding, we guided participants to complete an income form for one of their recently completed tasks (e.g., an Uber trip, a Rover Task, or an Upwork job) while they screenshared. For the remainder of the session, we introduced the rest of the features of Gig2Gether. Participant's screenshares and real-time interaction allowed for immediate feedback and clarification. 
At the end of each session, we detailed the study's minimum requirements and optional tasks -- a copy of consent form, which includes the payment structure, was also sent to each participant via email. The daily task requirement rewards participants \$15 a day for completing one of:
\begin{enumerate}
    \item Upload entries on \textbf{expenses} incurred (e.g., gas, pet supplies, office supplies) or \textbf{incomes} earned, which include
    \begin{enumerate}
        \item  Trip for Uber
        \item Income forms for Rover or Upwork
    \end{enumerate}
    \item Share a story
\end{enumerate}
To earn the bonus, participants were expected to complete the daily task for each of the 7 consecutive days, on top of completing the following actions at least once: 1) Plan upcoming work, 2) View \textit{Personal Trends}, and 3) Like another worker's story. 
Optionally, Uber drivers received the option of earning the bonus by uploading a CSV of historical trips, in lieu of the three actions listed above. 

\subsection{Exit Interviews} 
We conducted one-hour semi-structured exit interviews with each participant. Questions of the protocol focused on the key features of Gig2Gether—such as data uploading, trend analysis, storytelling, and the planning tool—as well as participants' overall experiences. Additionally, we tailored questions to the records of participants' daily interactions with Gig2Gether, including stories shared and uploaded income/expense entries. A complete list of exit interview questions is available in Appendix \ref{exits}.
\subsection{Analysis Method}
To investigate workers' interactions with our system, we took a mixed-methods approach to 1) aggregate quantitative statistics about usage such as counts of stories/uploads shared and 2) qualitatively examine onboarding and exit interviews. 
Quantitatively, we fetched and aggregated usage reports directly from the system backend, and then performed minimal calculations (e.g., counts, averages).
For interview transcripts, three researchers conducted open coding to identify concepts, themes, and events. 
\input{story_stats.tex}
\section{Findings}
Below we present the three main finding themes. First, we describe the role that Gig2Gether played in participants' workflows as well as (current and imagined future) use cases regarding the tool; next we give an account of workers' stance on the system as a means to share data with policymakers and the types of information they prioritized to share; finally, we introduce new considerations for a worker data-sharing system as surfaced from participants' use of Gig2Gether during the field study.

\subsection{Worker Data-sharing in Practice{: Exchange Support \& Manage Individual Finances}}
During onboarding, participants expressed initial reactions to how they envisioned using features of Gig2Gether. 
{In} exit interview{s after the 7-day field study}, workers {shared further }details about {existing and desired use cases.}
In the following, we present findings about how participants used features of Gig2Gether using contextual details they revealed during interviews and usage metrics gathered from the system.

\subsubsection{{Solidarity \& Collectivism via Experience \&} Data Exchange} 

Many workers described \textit{Stories} as a unique feature distinguishing Gig2Gether from other data-tracking or -sharing apps they use. Though not everyone shared, many workers found it reassuring to read others' stories, since they get to \textbf{learn that they're not alone in experiencing hurdles} at work: ``I really like the fact that there's stories, and you can check out what everybody else is dealing with. So you feel like: Oh, I guess it's not just me that's feeling like they're \dots be[ing] cheated'' (Freelancer-1). Freelancer-2 shared the desire of wanting to connect with others, because ``You can really feel siloed as a gig worker sometimes, so it's cool to see other people's experiences''. When first reacting to the story feed during onboarding, Petsitter-5 immediately expressed resonance with a story: ``I have similar feedback \dots I'll be adding a story shortly, because it's hard to get [jobs on Rover] versus \dots WAG. Yeah, definitely want to talk about this.''
A few workers specifically pointed out the content and attitudinal contrast of Gig2Gether's story feed with other gig work forums: ``the subreddit is really just a lot of sharing \dots but not necessarily useful [content] \dots but [Gig2Gether] offers tools'' (Freelancer-2). One participant even expressed considered sharing stories to initiate collective action against Uber:

\begin{quote}
There's a lot of things that I would like to share, but most of them are political. So like: we should all get together, fight back against Uber \dots [but] I didn't know how political I could be [on the Story feed].
\end{quote}

Several participants shared a displayed level of \textbf{interest in other platforms} supported by {Gig2Gether}---either they had prior interest or developed interest for how to start work on another platform after reading others' stories. In both cases, workers found value in reading about others' experiential strategies and issues. {This interest in other platforms' users' stories was reflected in usage metrics (see Table \ref{story_stats}), which show how platforms' workers expressed support (via likes) for a comparable number of stories in their own domain as from other domains (likes from other platforms are bolded).} Petsitter-3 is now considering both Uber and Upwork as extra sources of income: ``I did [like] one [Story] from Upwork because I was actually looking to work there for Upwork at some point \dots I saw a lot of pointers that people gave for Upwork, and I was like, `You know what? I'm gonna definitely keep that in mind.''' 

\paragraph{\textbf{{Collaborative Examinations of Algorithmic Speculations \& Rate Standardization}}} Although the \textit{Collective Insights} page was not yet populated with real user data, it sparked ideas and hope in participants for what could be revealed with aggregated data. For instance, Driver-2 expressed excitement about the potential of \textbf{answering popular speculations} about effects of having a Tesla on Uber earnings: ``[on] the Reddit Forums for Uber drivers, people are always asking `if I buy a Tesla (or if I get an XL) what should I expect as far as [how much] my tipping [were] to increase, or hourly income to increase?' So this is actually pretty cool''. In addition to large differences such as car model, Driver-7 wondered whether small gestures such as amenities can affect earnings: ``car model \dots [and] the type of amenities that the driver offers''. On Rover, Petsitter-2 also wished to confirm her own observation-based hypothesis that ``vets have a lot more repeating customers \dots they also tend to be the more expensive ones''. Beyond helping workers decide the type of services to offer, participants also saw collective insights as a tool to help them \textbf{set rates of charge for services}. Petsitter-4 expressed how
\begin{quote}
    I would love to see [earning statistics] broken out by urban, suburban, rural \dots [because] that's the biggest difference in how sitters operate \dots it's a entirely different game. Right now I'm urban, I have a radius of two miles and I walk to all of my bookings, whereas a rural sitter might have a radius of like 10 miles, where they'd have substantial costs in terms of travel time and driving \dots [So urbanization would impact how] I set my pay rate. 
\end{quote}
Driver-3 similarly wondered about fare price difference across geographic regions: ``The only [additional breakdown I'd want] \dots would be your region \dots I noticed different fare prices of getting out of the city''.

In online freelancing, platforms offer a wide variety of job categories, thus Freelancer-2 desired to find out about differences between and intersections of categories: ``I work in healthcare but a lot of the work I do on Upwork is writing, it would be interesting to see \dots [the breakdown or] an overlap of both categories.'' Freelancer-2 {also} offered the idea of breakdowns by disabilities: ``physical and mental disability, might also be a good differentiator there''.

\subsubsection{ Integrated {Financial} Tracking{: From Self-Logging \& Reflections $\rightarrow$ Planning \& Managing Accountabilities}}

\paragraph{\textbf{Streamlined Financial Tracking}} Participants described Gig2Gether as straightforward (``simple'' and ``easy'') to use when manually entering information. 
While various third party apps emerged over the years to help workers track earnings, expenses, and tax obligations --- as noted in \S\ref{Related_Work} and by workers such as Driver-1 -- some participants (particularly from {non-driving domains}) preferred Gig2Gether over such apps for its simplicity: ``I like this way better, because this is for gig workers, and the other is more of the financial crap that I don't like having to deal with, but I do [have to]'' (Freelancer-1). Petsitter-5 also enjoyed the simplified experience of viewing his financial data: ``this is better than [Rover. There] it's just too complicated. And I love seeing how data is simplified [here]''.

Although rideshare driving often accumulates a larger volume ``gigs'' in a day than petsitting or freelancing, D3 (who did not previously use tracking tools) also expressed a desire to using Gig2Gether: ``I don't always remember everything, but I can keep it all just between the Uber and my head. But I'd like to use a simplified [tracker]--- another platform like you guys are presenting now.'' 
\input{summary_stats}
\paragraph{\textbf{{Integrating Financial Reflections into Gig Workflows}}}
Beyond the initial income entries uploaded during onboarding, all participants entered additional income entries, and 9 of 16 uploaded expense entries {-- with one participant entering 7 expenses (see Table \ref{summary_stats})}. Based on these uploads, participants reflected on the personal earnings presented by the Trends page. Driver-2 appreciated having the ability to review his work data: ``I really liked the Trends section. Uber doesn’t give trends, just reports. And the Trends helped to look back and choose the weekends and decide what times are best to work.'' During onboarding, (part-time) Petsitter-3 expressed similar excitement about being able to compare earnings across time: ``Rover doesn't have something like this where you can \dots compare this year to last year.'' 
Driver-6 was excited by the ability to view his weekly data{, and} wanted to use the Trends page to show his friends proof of earnings having gone down, e.g., from working the same amount of time, year over year:
\begin{quote}
    If I have this app, then I can show them the facts \dots this is this [amount] before, and this is now \dots it can actually affect if I still want to do this Uber thing, or I can tell my friends not to do it anymore. \dots Because this is data, this is like facts 
\end{quote}
Driver-3 went a bit further to imagine how historical data can help him plan for breaks:
\begin{quote}
So you can cut out with Uber, cut out some downtime with the slow hours \dots actually have a break and not worry about missing anything. \dots [I could] look back to last year and say \dots September 1 was busy, and it slowed down at 10 o'clock \dots So you don't have to waste your time staying on the app    
\end{quote}

Driver-3 liked recording and seeing expenses displayed back, explaining how Trends could help him ``streamline my expenses a little better \dots when I use plastic [cards to pay], I don't pay attention as much as you do when you're handing cash over \dots [but] with having your site up, I could just go back and refer to everything and say, `Hey, maybe slow down on this' \dots when I'm going through expenses.''
In a similar manner, Driver-6 enjoyed entering his information at the end of a day that he had driven. In his 10 years of driving, he had never been compelled to try recommended apps from fellow drivers, but found it easy to use Gig2Gether to enter information and subsequently view Trends.

Freelancer-1 also shared the enthusiasm for potentially streamlining work processes such as tax filings: ``This is super helpful. If [only] I would have had this when I was helping my husband with starting up his stuff \dots the whole tax thing was a nightmare for me''.
Petsitter-1 described her affinity for the tailored aspects of the tool by contrasting against how most similar apps frame gig workers as independent contractors, which misaligns with the reality of their work and earnings: ``A lot of that stuff is like: `Get \textit{blah, blah blah} for your small business.' [But] I'm not a small business owner''. 
Several participants talked about creating reminders to remember to upload their data, such as Freelancer-2 ``a reminder in my calendar just to make sure I wouldn't forget'' and Driver-3, who had to ``set a reminder to make sure I did [uploads for daily tasks]''.
However, uploads can became a part of normal work routine---Petsitter-3 added it to the ``housekeeping things I needed to do, and it seemed to flow pretty naturally in with those reminders.'' In the same manner, Petsitter-4 also mentioned push notifications would help but were not necessary because ``anytime you start something new, it's not habit yet \dots just have to get used to it''.
\vspace{\baselineskip}
\paragraph{\textbf{{Planning, Keeping up with \& Achieving Earning Goals}}} Although the Gig2Gether \textit{Worker Planner} was only populated with mock data, participants were eager to incorporate it into their workflow\replace{. 
Workers}{, and} resonated with the planner's goal of helping them structure schedules for days and review earnings goals. Freelancer participants foresee themselves using the \textit{Planner} to track true hourly wages after expenses: Freelancer-2 would use it to``keep things straight \dots [So I can compare:] I work this much this week. This is how much I uploaded [in earnings]'', while Freelancer-1 would use it to check ``how much I'm working, whether my expenses offset with the money I'm making. And see if I need to work more''. Driver-3 envisioned using the \textit{Planner} to help remember and plan around upcoming reservations, which can go as far out as 30 days:
``I would definitely use it a lot, because of the reservation rides \dots tonight I have a reservation ride for [which] I can't remember [the exact time]''. Petsitters held mixed opinions about the \textit{Planner}, partially due to how Rover already provides a calender for bookings --- we outline some suggestions they made in \S \ref{planner_improvements}.

\input{tags.tex}

\subsection{Data Disclosures to Policymakers, Peers \& Advocates}\label{Findings_Sharing}
Workers also envisioned potential ways of impacting policy or mobilizing collective action for several initiatives, described below.
\subsubsection{Openness to Data Sharing with Policymakers and Advocates}
Workers expressed strong support for Gig2Gether's mission of shedding light on their working conditions to policymakers: ``This is a tool that's designed to bring exposure to policy makers \dots \textbf{to open the door between drivers and politicians} \dots now that could interest a lot of drivers right there.'' (Driver-1). 
{Through shared stories, we note workers were willing to share their qualitative data with policymakers in 23 of 27 cases (Table \ref{story_stats}). }
Beyond a willingness to share data with policymakers, workers also shared preferences for prioritized issues such as safety and wage concerns. With Gig2Gether, they hoped advocates and policymakers will ``get out the realistic facts of the jobs'' (Petsitter-1). Even when they were not sure how a story or metric could relate to policymaking, participants exhibited a general desire for their data to simply raise awareness about their work conditions: ``You could share that [data] because I don't think anything would hurt anyone. If anything, it'll maybe open some eyes up.'' (Driver-3). 

\subsubsection{{{Story Feed: A More Reputable Source for Informing Policymaking}}}
When comparing the \textit{Story} feed to other online forums they engaged with, participants considered Gig2Gether as a more credible source, which may make (1) policy stakeholders take it more seriously, and (2) workers more comfortable interacting with other workers. Petsitter-4 explained her rationale for increased trust in Gig2Gether: 
\begin{quote}
I would feel a little bit more comfortable that I was getting information from like verified sitters \dots
it would be weird for someone to sign up for an app to track their earnings, and then shit post in the community section of it \dots It would be a community that would be obviously a little bit more verified, and then a little bit more serious \dots [with members who are] committed to gig work, to a point where you're going to the trouble of tracking your earnings/expenses''.     
\end{quote}
Sharing this sentiment of increased reputation/trust in Gig2Gether, Freelancer-2 postulated on its effects on policymaker perceptions: ``I feel like \dots they might disregard what they saw on Reddit \dots [as] people venting online, people being bitter\dots but if it was coming from a more reputable forum \dots They might take it to heart a little bit more.'' Driver-2 also compared it to Reddit, saying Gig2Gether represents a place with less trolls, where workers are ``planning for more success''.
Meanwhile, Petsitter-1 shared her thoughts about the role of advocates in disseminating information about the platform: ``[Gig2Gether and its insights] is the kind of thing that I think would be better spread through advocacy groups than through individual word of mouth ''

{Using} the \textit{Story} feed, workers shared more strategies than issues, with ``safety'' and ``{fair pay}'' emerging as the most used tags. 
For strategies, many workers talked about staying safe in the face of challenging client interactions (for Rover users, ``client'' can refer to the pet and/or its owner), such as Petsitter-1 when watching multiple dogs and Driver-1 when faced with trespassing customers---see Figure \ref{phone_customer}. 
Experienced workers also shared strategies for improving earnings, such as methods for attracting repeat customers (Petsitter 2, Figure \ref{p2_repeat}), testing platform features (Driver-2, Figures \ref{reservation}, \ref{d2_airport}), recording unpaid work/time (Petsitter-4, Figure \ref{p4_unpaid}), or even a workaround for platform's evasions of small fees --- by tracking them and filing small claims lawsuits (Driver-1, Figure \ref{small-claims}). 

In terms of issues, workers most commonly shared experiences of unsafe working conditions --- e.g., a driver writing about a stressful trip taking a distressed elderly man to the ER, Figure \ref{d2_hospital}. 
Workers also shared frustrations about understanding how algorithms assign work (Drivers-3,7 in Figures \ref{d3_power}, \ref{d7_platform}) and concerns of power imbalances with clients (Petsitter-2, Freelancer-1 in Figures \ref{power}, \ref{f1_power}).

\subsubsection{{Safety \& Wages}} 

{During exit interviews, we probed workers about how and which these shared concerns should be communicated to policymakers.} Below we detail examples of compromises to their safety or pay. 

\paragraph{\textbf{Understanding Worker Safety}}
When discussing safety concerns, participants referred to \textbf{physical safety} issues they face from riders (Uber) and pets and/or their owners (Rover), as well as \textbf{digital scams} (all platforms). We describe below the various physical dangers, but expand on scams in more detail in \nameref{scams}, since participants did not prioritize these as a concern to share with policymakers.

Driver-1 described various factors that drivers might encounter ``incidents \dots like physical assault, being disorderly, and causing damage to the driver's vehicle (this happens pretty often), passengers getting arrested out of the back of your car'', which motivates him to use a channel such as Gig2Gether's \textit{Story} feed to funnel the information to policymakers, since it ``would be good to be able to report that somewhere centralized so that they can see there's a big safety issue.''
Beyond road conditions, safety risks can also encountered at strangers' homes ``you're going into somebody's house, it's a vulnerable position to provide work'' (Petsitter-4). Furthermore, both participants pointed out how many of the risks imposed on workers are one-sided to protect consumer identify and safety: ``sitters are background checked, clients are not'' (Petsitter-4), and Driver-9 related being required to pass``a pretty rigorous background check \dots Both initially, and then it happens randomly. Usually only once a year, but it \dots has been more often''. Driver-1 described how prior to the \#WHATSMYNAME movement, ``the passenger would give their name to the drivers so that the driver knows that they have the right person'', but nowadays drivers have no method of verifying whether they have the correct person, causing breaches of safety because 
\begin{quote}
    You got young, beautiful women in their early twenties out there driving, and some big, burly dude opens the door [and asks] `What's my name?' Whatever name she puts out, he could say yes, [and] she could disappear from that point.
\end{quote}

\paragraph{\textbf{Understanding Unfair or Unpaid Wages}}
Participants of the three platforms described scenarios related to unfair or unpaid wages. Senior Drivers-1, -6, and -7 all lamented how Uber wages and incentives keep dropping over the years: ``when Uber first started, we were making like almost \$40 an hour. Now it keeps on going down \dots [on] the Quests right now, you just make make \$15 on 20 trips \dots they're getting so greedy'' (Driver-6) and ``Uber has gotten worse, and this might be my last summer [with them]''. For many of these rideshare drivers, gaining access to collective evidence is quintiessential for exposing the rapidly wage declines: ``The reason why \textbf{this data is important is because we want to expose literally what we're making. We want these policymakers to see this}''. Even more junior workers such as Driver-10 expressed desires to use the app to record subminimal wages: ``some states \dots looked at it and said, this [wage] is not fair. So I think that's probably where I would use the data that's within your app to basically show, `hey, here's what's really going on`. ''

\subsection{New Data, Metrics and Features}\label{Findings_Improvement}
{While research has explored workers' preferences for contributing data and how data can be used \cite{stein2023you, supporting}, having workers test a prototype can allow them to surface important needs and opportunities that only arise from hands-on experience \cite{rogers2007s}. We found this to be the case where participants' use of Gig2Gether revealed important workflows to support, opportunities of insights to strengthen personal goals and collective action, as well as worker-to-worker interaction and anonymity preferences that would have otherwise remained unknown.} 

\subsubsection{{Insights About Essential Workflows to Support in Data-Sharing Systems}}

{As participants described their experiences using Gig2Gether to log their work, they highlighted additional important workflows that must supported for them to obtain the most useful insights about their work and earnings.}

\paragraph{\textbf{{Towards} Complete \& Automatic Data Uploads}} 
Several participants talked about taking gigs off-app (Petsitter-3, Freelancer-2) or {working} multiple apps (Drivers-1, 6, 7). For example, Driver-7 has shifted ``90\%'' of his work to Lyft so the Trends page would not reflect all his gig work earnings. He currently uses an Excel spreadsheet to manually input his weekly summaries from both platforms but wants an app that helps him track both.
{By describing their experiences uploading data and viewing their trends, participants highlighted the importance that} future versions of Gig2Gether {support workflows of multiple apps and off-app work so the} Trends page {allows them} a holistic {and meaning} view of net earnings and patterns. 

{Additionally,} related to data completeness, several Uber drivers described a need for automatic data entry support (Drivers-1,7, 9), similar to existing third party apps (most of which require paid subscriptions) that emulate actions on Uber/Lyft such as accepting or declining ride offers \footnote{Examples include Mystro and Para, both of which are paid apps}. Especially for full-time and long-tenured drivers, the volume of trips they accumulate can be substantial. Even though we offer a CSV upload option {for Uber drivers} to mitigate the process, drivers {describing the many trips they complete a day and the normalcy of switching between apps highlighted the importance of automatically gathering their work data to support complete data insights}. {When asked about concerns around data privacy if their accounts were linked, drivers did not have any and were supportive of a more automatic option.} {On the other hand, we anticipated that Petsitters or Freelancers would not require an option of bulk data upload given the nature of their work, and correspondingly, they did not share a need for automatic data entry.}

\paragraph{\textbf{A One-Stop Shop {For Understanding Profit \& Filing Taxes}}}
{Participants' descriptions of their experiences entering completed work and accumulated expenses also helped us recognize a possibly under-supported task in the ecosystem of gig worker tools: the tax-filing process.}
Driver-6 and Petsitter-1 {described desiring a ``one-stop shop'', which for them} translates into {one tool that lets them} pull all their data for purposes such as submitting to an accountant, IRS audit, tax filing, or if they're just curious:
``a one stop shop [so] that I can see my progress. I can see how much I'm making per hour. I can see my expenses. At the same time that I can show it to my accountant. Or if there's an audit from IRS, that I can show this.'' (Driver-6). Mileage, in particular, {was highlighted by both drivers and petsitters as an important metric for accurately calculating their expenses for filing taxes under the standard deduction:} ``I would like \dots [for this app to have] as opposed to them [other apps/forums] `one stop shop', if it had the mileage.'' (Petsitter-4).

{A couple participants described using a combination of apps to collect all the metrics related to understanding their work and filing their taxes.} Driver-7 uses {Stride to automatically track miles and Excel to manually log trip earnings.} 
{He expressed that Gig2Gether automatically} tracking miles would {complete the metrics} he needed {in a tool}---mileage, earnings and expenses altogether. 
{Driver-9 also shared using multiple apps for tracking mileage (Gridwise) and fuel expenses (Fuelly). He explained that Gig2Gether tracking miles} would further motivate him to share the app with friends ``I would recommend it, because it's more immersive than the other app that I use [especially if it can also be used] to track your mileage''.

{Not all participants desired this though.} Driver-8 warned us against trying to {expand Gig2Gether's features to fulfill} a ``one-stop shop'', {expressing worry around the} risks of chasing down an {endless} pool of desired features. Instead, he encouraged us to pursue Gig2Gether as a tool for connecting workers with policymakers and advocates as this was the unique feature he had not seen in past applications.

\paragraph{\textbf{Providing Context on the Planner}} \label{planner_improvements}
{While the \textit{Planner} was primarily presented in Gig2Gether as a predictive tool to project weekly earnings, participants offered different ideas for how they wanted to use it by describing their current work planning process.}
{For instance, though Uber's traditional model has been on-demand ride requests, they began letting riders schedule a ride request in advance---``Reservations''. Driver-3 actively accepts these trips and wanted} to use the planner to keep him accountable for reservations. {Meanwhile} Petsitter-2 {wanted to use the \textit{Planner} to keep track of the different pets she's booked to care for}:
``Say I got Ice or I got Henny \dots I put their names all throughout the planner \dots Because sometimes I get them back to back and it'll be like: `Okay, wait, who's this one? ' '' Petsitter-3 also entertained ``the possibility to write in who I'm pet sitting for'' and further suggested the idea of ``being able to put in the address'' to each entry.

{Several drivers also highlighted their tendency to center their driving locations and hours based on events. Thus, they} pointed out the utility of incorporating regional ``events that would be in the city'', but not bigger ones because Uber already keeps track of those. Driver-7 {clarified how seeing events integrated into the \textit{Planner} would help smooth out his current workflow, as right now} he resorts to manually looking up such events himself, which can be time-consuming: ``I have to go online \dots [to look up when], Chicago Cubs play \dots write down on calendar by hand the right time \dots''.

\subsubsection{{New Metrics to Strengthen Personal Goals \& Collective Action}}
\paragraph{\textbf{Net Earnings Insights: Achieving Personal Goals and Empowering Collective Action}} 

{Reviewing quantitative metrics on Gig2Gether, participants talked about how these can help support their personal goals as well as ideas they have for advancing collective action.}
Uber and Upwork participants yearned to view their net earnings (Freelancer-1, Driver-1), so they can plan for and achieve work goals. For instance, Freelancer-1 wanted to ``see how much I'm working with my expenses offset with how much money I'm making, and see if I need to work more''. Beyond personal earning goals, Driver-7 wanted to leverage trends from earnings data to show other drivers unfair or demanding working conditions imposed by platforms, explaining how workers often focus only on gross income without critically assessing their expenses. For instance, he'd want a way to show drivers whose net earnings are below minimum wage---e.g., 15 hours to make \$200. These statistics ``give them an insight of what's going on \dots [that] they're not making enough money'', so as to galvanize them to strike against rideshare companies, because ``in order to make a change, we [as drivers] have to get together'' in protest.

\paragraph{\textbf{``Downtime'' and ``Deadtime'': Visibilizng Total Work Time}}
{Participants also suggested additional metrics to improve workers' understanding about time they spend working that they might sometimes overlook.}
To optimize working time, they explained the importance of including metrics and visualizations that illustrate not only of hours {actively booked on a job}, but also hours that are unaccounted for, such as ``downtime''---i.e., time spent waiting for work opportunities (Petsitter-4). 

Drivers also wanted to record and view ``downtime'' --- or ``dead time'', in their case, referring to time spent driving without a paying customer in the car.
Drivers-1 and -7 both deemed it imperative for drivers to know proportions of their paid time within total working time --- which should include paid trips, time spent driving towards customer pick-up, and wait time between trip requests.

\begin{figure}[h!]
    \centering
\includegraphics[width=\linewidth]{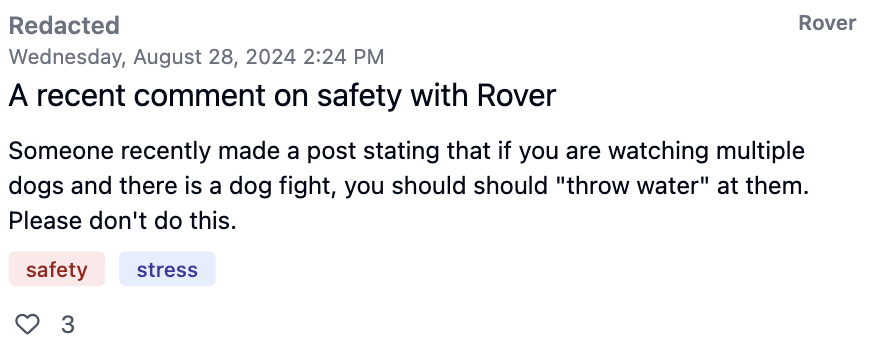}
    \caption{Petsitter-1's response to another strategy}
    \label{dogs}
\end{figure}

\subsubsection{{Additional Worker Interactions \& Anonymity Preferences}}

\paragraph{\textbf{Commenting \& Reaction Options}}
Currently, Stories have limited interactions: workers can post a story, read a story, or like a story. Participants held mixed feelings about new interactions: Petsitter-1 was adamant against implementing additional interactive (commenting) features ---``I don't think it's productive.'' --- but many others expressed desires for comments and moderation mechanisms (Freelancer-1, Petsitter-3, Driver-2, Petsitter-5). Ironically, Petsitter-1 did make a new post to respond to another sitter during the study Figure \ref{dogs}. Petsitter-4 {asked} about capabilities for networking with others--- \replace{she did this previously and finds }{based on previous experience, she found }it helpful to be on a list of trusted contacts that refer each other to clients. 

Yet, others wanted to prevent the Stories feed from duplicating behaviors of forums (e.g., subreddits or Facebook groups). Participants especially wanted to avoid off-topic posts or misunderstandings --- ``a lot of stuff gets taken out of context'' (Driver-3). 
Petsitter-4 suggested adding ``Agree'' and ``Disagree'' buttons; Driver-1 proposed allowing comments, but then only giving viewing access to policymakers, so as to minimize worker disagreements: ``let's say when a `Driver A' posts a story \dots I've dealt with the same issue, I comment on that story. The other drivers would not be able to see my comment. But the policymakers would be able to, which means no online arguments amongst drivers''.

Driver-6 gave a unique idea to merge qualitative and quantitative data, so as to improve readers' confidence in the veracity of stories being shared. He uploaded Trip data and described an issue he faced with Upfront fare in the Notes field, and wanted to share this data with workers, policymakers, and advocates within the Story feed. He felt stories with real Trips attached could serve as proof or evidence that a story is not fabricated. On the other hand, Petsitter-5 did not believe pay information should be shared, but did encourage more highlighting of photo and video sharing options.

\paragraph{\textbf{Anonymity and An ``Edited'' Trail.}} 

Several participants requested more anonymity and traceability on the \textit{Story} feed, especially necessary once Gig2Gether is circulated and used more widely by different stakeholders. Though many participants were comfortable sharing stories with their usernames, others wanted strict anonymity, pointing out that they and others may use the same social media username across platforms. This could uncomfortably lead to being identified if peers of subreddit or Facebook groups join the platform and disagree with their stories (Petsitter-4). Driver-7 also wanted to share ``political'' stories but refrained because there was no way to post anonymously.

Drivers-1 and 6 both emphasized the importance of removing location information as well. Driver-1 asked: ``Is there any situation that would cause the trip data to be visible to other drivers? For instance, would they be able to see what city these are in at any time? \dots That's a very, very serious concern.'' Driver-6 advised for Gig2Gether to automatically blur sensitive information, such as addresses, that workers might upload, explaining drivers often forget this step when posting screenshots on forums which risks privacy and security. Driver-1 also highlighted the need to maintain an edit trail for posts. Users have different reasons for editing stories---a few participants said they appreciated the ability to go back and modify typos. However, he was concerned if deceitful users were to abuse the current lack of an edited label to gaslight others.
This feedback underscored the importance of ensuring robust privacy controls within the application. 

\section{Discussion}

\final{Our study extends prior work on policy-related stakeholders' ideas about role of worker data in policymaking---e.g., aggregated data for investigating unfair pay or discrimination \cite{supporting}, formats of data probes to interactively illustrate workers' platform experiences to colleagues \cite{zhang2024data}---with complementary worker insights.}\del{Literature has surfaced what policy stakeholders feel gig worker data can be useful for---e.g., drafting bill language, communicating workers' platform experiences to colleagues \cite{zhang2024data}. Research has also reported several initiatives and related data metrics that policy stakeholders desire for supporting worker-centered initiatives---e.g., aggregate data to investigate unfair pay or discrimination \cite{supporting}. Our study extends these insights with workers' initial impressions from using a data-sharing system intended to support individual goals and inform policymakers.} 
First, we offer implications for how Gig2Gether could assist workers in supporting policy for \textit{worker safety} based on the stories participants shared. Then, we discuss challenges for building a data-sharing system that ensures anonymous and truthful sharing. Finally, we reflect over how data-sharing on its own should not be conflated as a catch-all solution by imagining how a system like Gig2Gether can complement worker empowerment and collective action efforts.

\subsection{Gig2Gether: ``A Door Between [Workers] and Politicians''} \label{door}

Participants exhibited a weariness about whether actual policy or regulation \final{is achievable}\del{would pass to improve their working conditions}: ``Except for you guys, no one is trying to help us \dots no one is trying to expose any of the issues \dots If there's an issue with a passenger, the politicians are all over it. But with a driver, [not so much]'' (Driver-1). This skepticism for change is a natural barrier to the system reaching a critical mass of users before it can gain momentum in pushing forward policy initiatives.

However, participants were also \textbf{roused by \del{the promise} Gig2Gether \del{represents} as a vehicle \del{for overcoming systematic challenges,} to enable policymakers and advocacy groups access to data and issues workers face} and advance \del{change on }labor {protections and initiatives}. \final{For example,} recall how Driver-7 suggested a new tag, ``Political,'' so workers can prioritize which stories policymakers should see first. And though participants conveyed discouragement that policymakers are more invested in addressing consumer protections than worker issues, we observed several discussed platform issues by explaining the negative impact on \textbf{both} workers and customers. For example, Petsitter-4 shared that Rover appears to be double charging fees on customers, and Driver-1 described how Uber's current verification policies puts drivers' and customers' safety at risk. \footnote{{We did not have a chance to ask if tying together worker and customer issues was unintentional or motivated by the belief this would garner strong policymaker attention.}}
{Though just one week, our field study is a nascent glimpse into how several participants used the platform to contribute policy-aligned narratives and suggestions. This makes us wonder whether continued contributions to a data-sharing system can 1) gradually influence workers not usually inclined to participate in ``political'' initiatives, to re-frame their motivations for data contribution, and 2) foster cross-stakeholder collaborations for actionable policy goals.}

\del{Reflecting on }Workers' Stories also suggest to us Gig2Gether's potential \del{as a mechanism }to influence policy for gig worker safety---the top concern shared in participants' Stories---in meaningful ways. Most gig worker legislation and regulations in the U.S. address minimum pay standards (e.g., NYC, Chicago, CA, WA, MN), and more recently data transparency (e.g., CO). Yet little exists for worker safety. Colorado's recent data transparency bill contains \final{limited} language on delivery worker safety, \final{only} require platforms to send the customer a nudge encouraging them ``to ensure driver safety upon arrival, including by ensuring a clear, well-lit, safe delivery path and ensuring all pets are properly secured.'' \cite{Colorado_General_Assembly_dnc_2024}. 
A more \final{purposeful} attempt is NYC's regulations for food delivery workers' physical health and safety: the text explicitly grants workers access to restaurant bathrooms and allows them to ``set limits on travel over bridges or through tunnels and the distance between a restaurant and a customer'' \cite{NYCFoodDeliveryLaws}.
Participants' stories on Gig2Gether were rich and \textit{specific}, describing customers trespassing property (Driver-1), experiences breaking up animal fights (Petsitter-1), clients violating Terms of Service (Freelancer-1). These are all serious scenarios that workers were unsurprised by\del{, given their own experiences,} but are likely non-obvious for policymakers or the public at large. Taking inspiration from \del{the example set by }NYC's delivery worker protections, we propose workers' stories on Gig2Gether could be leveraged \del{in bill writing to} for bill language and requirements that meaningfully protect workers' physical and mental well-being.

\subsection{Ongoing Challenges of Stakeholder Verification: Ensuring Anonymous, yet Truthful Sharing} \label{ongoing}

A data-sharing system that provides collective insights for different stakeholder groups must ensure both worker privacy \del{(i.e., their data and identity in case of}to protect workers from platform retaliation, and data reliability \del{(i.e., that the contributors are real workers and the data is }\final{to verify that contributions are} true and complete). Our pilots and field study were invite-only, allowing us to limit access to only \del{participants verified as qualified}verified workers for our study. However, \del{even with this closed system, }we faced challenges of securing a data-sharing platform \del{faced by Gig2Gether and advocates or grass-roots organizations}\final{that future efforts} pursuing a mission of data-driven insights towards worker well-being \final{should keep in mind}. 

First, while building the system, we were unable to find platform APIs to link gig worker accounts and validate both worker identity and completeness of their platform data uploads. As noted in \citet{dubal2023algorithmic}, third party data connectivity services exist \footnote{\href{https://argyle.com/}{argyle.com/}}, but are expensive and have questionable data privacy and protection practices. Additionally, as our participants raised, manual entry would still be necessary for workers who accept off-app work. As discussed by \citet{supporting}, workers may take work off-app to \final{supplement} low \final{platform} wages, \del{which would be important} critical information that can help policymakers determine whether gig platforms are creating untenable wage conditions. 
During field evaluations, we requested workers to share their in-app profiles during the onboarding  Zoom session for verification, but this process is not scalable. Outside of APIs or data portability services, there are few fool-proof document-based methods to verify workers \footnote{Several Upwork subreddit threads demonstrates the ongoing challenge clients experience in trying to verify freelancers to hire}. Finally, Uber drivers \final{sharing} the \final{(difficult to fabricate)} CSV files that requested from the platform \final{offered} a more reliable \final{verification} method\del{because it would be difficult to fabricate such data}. However, the other platforms do not accommodate such worker requests of data access, and we did not want to put Uber participants at risk of deactivation by requiring them to upload this file as a pre-requisite for study participation. 
\del{This leads to a related question \del{for publicly deployed systems}around the veracity of information.} 
These challenges \del{of affordable and accessible methods of} \final{around} data portability kept participants from easily connecting or uploading their work data from different platforms, and will \final{also} inevitably affect the quality of (and workers' trust around) collective insights and stories shared through the system.
\del{One potential way of} To mitigate this for the Stories feed, we consider Driver-6's suggestion of coupling Stories with related task-level data (e.g., a trip) to offer supporting evidence. 

{A caveat for data-sharing tools is \citet{khovanskaya2020bottom}'s warning that creating new data tools, especially ones to be managed by a union, can potentially burden union staff rather than enable change. Indeed, despite how workers entrusted us (researchers) with data, and Petsitter-1's idea for advocates as intermediaries to \final{raise awareness of} a data-sharing \final{system} to more workers, it remains unclear how to address issues of ownership. 
Harkening back to a collaborative model proposed by \citet{supporting}, one could imagine involving researchers \final{with} \del{management of}technical maintenance of tools but using them in collaboration with unions. For instance, researchers might explore stories and metrics alongside union representatives and workers to create membership recruitment material they desire for collective action.
}
\subsection{{Data-sharing to Complement Alternative Worker Empowerment Methods}}\label{complement}
{\del{We note that relying on }Data and policy as a standalone, catch-all solution should be cautioned against. We encourage researchers to consider creative alternatives and worker-driven objectives that would benefit if combined with data-sharing. We share two suggestions motivated by \del{the different ways}how participants used or imagined using Gig2Gether, as well as prior research on worker empowerment and collective action.}

{
\subsubsection{Informal Support Networks \& Mutual Aid.} Past work highlights the strengths of gig workers creating informal networks for mutual aid around purposes of companionship \cite{qadri2021s, atom}, pooling financial resources \cite{gray2019ghost, seetharaman2021delivery}, as well as sensemaking and strategy sharing \cite{mohlmannn2023algorithm}. These efforts can and do exist outside of a data-sharing system, and we do not wish to overlook other forms of assistance by overemphasizing a solutionist notion of data. Uniquely, participants' use of and ideas for Gig2Gether’s Stories feed suggests one way to complement worker mutual aid. 

First, Petsitter-4's inquiry about networking with other petsitters on Gig2Gether reminds us that not all gig workers have built-in communities \del{to lean on}. Public online forums like subreddits have low barriers to entry for seeking peers, but provide limited social connection --- Reddit users are anonymous and \final{having an account does not require identity} verification. Meanwhile, mediums like WhatsApp groups and co-located gig workers can establish intimate connections, helping build trust among workers for sharing mutual aid, but joining a group or \final{forming one's own} can be challenging. 

We recall that participants found solidarity and reassurance in reading others' \textit{Stories}, with some contrasting it as more productive and trustworthy than other forums due to its verified and ``more serious'' users. \final{Perhaps} data-sharing systems like Gig2Gether could offer workers a low barrier alternative to Reddit that \textit{does} allow for verification. This would help workers feel more comfortable and trusting of one another more quickly (akin to local Whatsapp Groups), an important \final{foundation} for successful social bonding. In this way, systems like Gig2Gether can be leveraged to strengthen workers' abilities to build personalized networks for mutual aid.

\subsubsection{Boosting Membership for Worker-Organizations.}
In the U.S., \final{a number of union activities have been reported related to collective bargaining} (e.g., striking, calling for boycotts \cite{Yamat_2024, Robertson_2024, Reuters2024vw}) and policymaking (e.g., fighting against anti-union law, pushing for worker-centered laws \cite{Quinlan_2024a}). In fact, the National Labor Relations Board (NLRB) report on recent data revealed that union petitions (i.e., requests to unionize) and support for unions have increased \cite{NLRB_2024}. Worker-organizing for gig workers has also gained traction, especially with the NLRB's 2023 reversion to a worker classification standard that offers gig workers a way to join unions \cite{Cockayne_2023}. 

It is unclear to what extent this has influenced gig workers \final{across platforms} joining unions \del{or the impact across different platforms}. Literature suggests that challenges in unionizing gig workers remain---\citet{schou2023divided} found that differences such as motivations and identities can lead to conflicting goals and hinder attempts at collectivizing, despite shared outrage over worker issues (e.g., wages). To counter those, worker-organizers have expressed interest in presenting workers with their data in formats like data probes to help them recognize platform manipulation \del{they perceive} and incense them into joining unions \cite{zhang2024data} --- echoing Driver-7's desire to use Gig2Gether to show workers collective insights about low wages to encourage a strike. Increased membership boosts unions' financial power to create change as member dues are crucial for unions to operate successfully---e.g., organize campaigns, negotiate and enforce contracts, provide training and legal assistance \cite{UnionCoded_2023}. 


}

\subsection{{Reflections, Limitations \& Future Work}}
{Atop the remaining challenges mentioned in \ref{ongoing}, we acknowledge several limitations that may constrain the validity of our study. First, our sample of three platforms restricts us from exploring other important avenues of gig work, such as food or grocery delivery. Relatedly, our sample of freelancers is disproportionate to the other two domains of physical gig work (due to issues encountered when verifying active and U.S.-based status), severely limiting our understanding of how remote gig workers respond to Gig2Gether.
While we did not engage with policymakers at this stage of evaluation, our results demonstrate active worker interest in collectively sharing data with policymakers to illustrate subpar working conditions.}
In the next phases of testing, we might consider interacting with relevant policymakers and advocacy groups to (1) evaluate the extent to which information gathered by Gig2Gether can affect policymaking and (2) collect additional resources and programs they might be aware of that can benefit workers, as well as potentials of open-sourcing the system.
{Finally, though our tool was designed to capture a diverse and heterogeneous set of perspectives, our reportings of findings in this paper may only capture a subset of more convergent opinions --- for purposes of forming a cohesive and concise narrative. We hope future (and related \cite{kuo2024policycraft}) works can examine ways of formulating worker narratives to inform policy \del{in a way that also highlights}inclusive of more pluralistic values, viewpoints, and cases.}
\final{It will be especially pertinent to consider how to carefully design data-sharing systems and policy interventions that support respectful, mutual understanding between workers of heterogeneous socioeconomic statuses, life stages, and cultural backgrounds.}

\section{Conclusions}
In this study, we contribute Gig2Gether, a tool developed with workers through thinkalouds, that offers them means for tracking financial well-being and exchanging information that can subsequently be shared with other stakeholders to affect advocacy efforts and policymaking. Through a field evaluation, we \replace{found workers to not only be willing and capable of using the tool --- they also generated ideas for new methods of collective data sharing.}{uncovered workers' existing and future use cases for Gig2Gether, in addition to ideas for new metrics to extend its potentials for collectivism and policy impact.}

\bibliographystyle{ACM-Reference-Format}
\bibliography{chi-references}
\input{appendixA.tex}

\input{appendixB}
\input{appendixC}
\end{document}

%% file: chi-related-work-2.1.tex
\label{Related_Work_Challenges}
In the U.S., \final{gig platforms and their workers} are often \final{referred to and regulated collectively}, where gig workers \final{(regardless of platform type)} are typically classified as independent contractors. This has resulted in limited policy or regulatory protections over work conditions, \final{burdening} workers \final{across platforms with confusing and overwhelming logistical} obligations related to self-employment{: navigating tax requirements \cite{taxing, tax, tax_lives, returns} through self-}tracking \final{of} earnings and expenses \cite{accountable, taming}, \final{conducting unpaid labor to find, procure and scope gigs \del{in times of precarity} \cite{youth, apouey2020gig, freelancecontrol}, assuming costs of work-induced injuries in lieu of workers' compensation and health insurance \cite{nilsenhealth, healthdrive, deliverysafe, hsieh2023co}, managing psychological costs to working alone \cite{atom, alienated, commodified}, and so on.} 

\final{Current research though has focused on highlighting issues specific to work context and platform worker-client matching methodology. For example, to the former,} studies surfaced safety hazards in ride-hailing and food delivery \cite{nilsenhealth, stressfulride, healthdrive, deliverysafe}, wage theft in care work contexts \cite{ming2024wage, cole2024wage, jerseycare}, or irregular schedules in online freelancing \cite{sousveillance, personal, freelancecontrol}. \final{To the latter, researchers have also reported how different platform mechanisms for matching workers to clients/projects impact worker autonomy/precarity.} \final{For example,} on-demand work \final{such as ridesharing and food delivery automatically matches workers and clients, limiting both groups' control in the process \cite{classification} and offloading} logistical overheads to workers \cite{own}. \final{And online marketplace platforms, common in freelance and caregiving contexts, employ rank-based methods and/or bidding mechanisms which skew client control over workers.} Ranking-based methods \final{allow clients to rank and sort candidates by historical performance metrics and relevance to projects \cite{context}}, pressuring workers to maintain \final{meticulous} portfolios \cite{making}. \final{Then, the bidding process where workers apply to postings \cite{personal}} requires additional unpaid labor from workers \final{to write proposals and scope projects without the promise of work} \cite{beyond, personal}.

Most of these studies examined issues with respect to a specific platform, thereby limiting insights into whether uncovered stressors generalize to other contexts \final{for strengthened implications of potential policy or regulation or more opportunities for worker community building}. However, in the few cases where multi-platform analyses were conducted, studies revealed how platforms \textit{do} present common higher-level risks (e.g., privacy, financial, psychological, gender biases) \cite{privacy, toward, brush}. \final{Not only that, but} the underlying causes of work challenges are often similar: the aforementioned lack of labor/safety standards and regulation gives way to unbridled worker exploitation through algorithmic management \cite{dubal2023algorithmic, machines, excessive}, gamification tactics and information asymmetries \cite{algorithmic, locus, zhang2022algorithmic}, and an absent collective worker voice that stifles public awareness of harms \cite{ming2024wage, cole2024wage, lastmile}. \final{These similarities in overarching causes of challenges to gig work suggest an opportunity for technology interventions to build solidarity between the currently fragmented and scattered worker communities.} Interventions for unifying gig workers \final{are a compelling exploration} since they often do not self-identify as gig workers, and instead use platform-specific terms (e.g., Uber driver) to self-describe \cite{supporting}, \final{weakening the potential for a collective worker voice to raise public awareness of individual \textit{and} shared harms} \cite{ming2024wage, cole2024wage, lastmile}.

%% file: chi_2.3.tex
\subsection{Gaps in Current Approaches to Gig Worker-Centered Data-Sharing for Policy}\label{Related_Work_Using_Data}
\final{
Recognizing \del{Increasingly, researchers and worker groups turn 
towards }the potential of work data to support workers' sensemaking and auditing of platform algorithms, researchers and worker groups increasingly turn to worker data exchange tools as a means to empower gig workers. Taking a first step towards gig worker-centered data collectives, \citet{stein2023you} deliberated on variants of data collection institutions with drivers, exploring \textit{data leverage} (also covered in \cite{levers}), \textit{governance structures}, and \textit{access control}.
Through lens of care ethics, \citet{sousveillance} used mock-ups to uncover freelancers' needs to relieve emotional strain, find legitimate gigs, and manage invisible labor -- emphasizing in particular how surveillance tools should not generate additional invisible labor for workers. 
To align worker needs with feasible policy changes, Hsieh and Zhang et al.
\cite{supporting} interviewed policy experts and co-designed with workers to understand their (mis)aligned policy priorities, in addition to the necessary supporting data needs. 
Although these studies surfaced key design requirements for worker data-sharing, the lack of a prototype working under realistic working conditions limits the degree to which they can identify and confirm the concrete challenges and practical needs of workers when integrating such hypothesized tools into their daily workflows.}

\final{Stepping beyond codesign, \citet{zhang2023stakeholder} created data probes to help Uber drivers explore and contextualize surfaced patterns with their positionality, well-being, and lived experiences. 
\citet{calacci2022bargaining} partnered with a worker organization to build the Shipt Calculator that audits algorithmic wage determination. Related organizations (e.g., Worker Info Exchange\footnote{\href{https://www.workerinfoexchange.org/}{workerinfoexchange.org/}} and Worker's Algorithm Observatory\footnote{\href{https://wao.cs.princeton.edu/}{wao.cs.princeton.edu/}}) formed to help platform workers collect data and investigate algorithmic decisions. 
\citet{imteyaz2024gigsense} used LLMs and online data (e.g., subreddits, app reviews) to help workers share and collaboratively identify issues, and subsequently generate solutions. 
While these tools leveraged data to support collectivism, few embedded into everyday gig workflows, and yet fewer considered broader, cross-platform gig work experiences.
Working with only quantitative wage data, the Shift Calculator limited understanding of worker struggles to one primary data type. While the seminal work of Turkopticon and Dynamo surfaced diverse labor issues related to policy changes
and helped workers form publics to mobilize/unify towards action, they focused narrowly on online crowd work, whereas challenges afflicting gig workers (especially those offline) diverge significantly. 
}

\final{
Finally, previous work explored how to leverage worker data to advance driver-centered policies. \citet{parrott2018earnings} performed a formative economic analysis of Uber/Lyft app data to investigate working conditions and wages of drivers in New York City, subsequently proposing a minimum wage standard that was adopted by the city. This data-driven approach to assessing driver minimum wage needs was replicated in Seattle \cite{reich2020minimum} and Massachusetts \cite{jacobs2021massachusetts}. Non-profits and other researchers followed suit on smaller scales with worker surveys, due to access restrictions to app data  \cite{Leverage_Dalal_2022, McCullough_Dolber_Scoggins_Muna_Treuhaft_2022, McCullough_Dolber_2021, washington2019delivering}. 
\citet{zhang2024data} explored how workers' data can support policymakers and policy informers, surfacing its potential to 1) inform policy creation, 2) support lobbying efforts, 3) support worker organization's (member) growth 4) aid regulatory efforts \footnote{One example is a nascent regulatory effort around \href{https://www.ftc.gov/business-guidance/blog/2024/03/price-fixing-algorithm-still-price-fixing}{algorithmic pricing investigations}}.
These efforts consisted of tools, systems or reports centered on quantitative data primarily pertaining to rideshare/delivery driving domains, but the policy changes proposed in such works focused narrowly on wages, missing insights and context on critical issues such as safety and discrimination. Lastly, such tools missed the opportunity to facilitate information exchange and collaboration between worker communities and supporting stakeholder groups, limiting their potential to align workers' community needs \cite[p. 4]{sousveillance} with policy advancements.
}

%% file: participants.tex
\begin{table*}[h!]
\begin{tabular}{|l|l|l|l|l|l|p{1.5cm}|p{1.5cm}|}
\hline
\textbf{ID} & \textbf{Age} & \textbf{Gender} & \textbf{Ethnicity} & \textbf{Tenure} & \textbf{Education} & \textbf{Household income} & \textbf{Gig Work Status} \\ \hline
Driver-1 & 45-54 & Male & White & 2-5 years & High school/equivalent & \$25-50k & Full-Time \\ \hline
Driver-2 & 45-54 & Male & White & 0.5-1 year & Bachelor's & \textgreater{}\$150k & Part-Time \\ \hline
Driver-3 & 45-54 & Male & White & 1-2 years & Some college, no degree & \$50-75k & Part-Time \\ \hline
Driver-4* & 45-54 & Male & White & \textgreater 5 years & Some college, no degree & \$25-50k & Full-Time \\ \hline
Driver-5\textsuperscript{+} & 35-44 & Male & Asian & 2-5 years & Professional degree & \textgreater{}\$150k & Part-Time \\ \hline
Driver-6 & 45-54 & Male & Asian & >10 years & Some college, no degree & \$25-50k & Part-Time  \\ \hline
Driver-7 & 25-34 & Male & Hispanic/Latino & 2-5 years & Bachelor's & \$50-75k & Part-Time \\ \hline
Driver-8 & 35-44 & Male & Asian & >5 years & High school/equivalent & \$25-50k & \text{Full-Time} \\ \hline
Driver-9 & 35-44 & Male & White & >5 years & Bachelor's & \$75-100k & Part-Time \\ \hline
Freelancer-1 & 45-54 & Female & White & \textless{}0.5 years & Associate's & \$25-50k & Part-Time \\ \hline
Freelancer-2 & 25-34 & Female & White & \textgreater 5 years & Professional degree & \$100 - 150k & Part-Time \\ \hline
Petsitter-1 & 35-44 & Female & White & \textgreater 5 years & Some college, no degree & \textless \$25k & Part-Time \\ \hline
Petsitter-2 & 18-24 & Female & White & 0.5-1 year & High school/equivalent & \textless \$25k & Part-Time \\ \hline
Petsitter-3 & 25-34 & Female & White & 2-5 years & High school/equivalent & \textless \$25k & Full-Time \\ \hline
Petsitter-4 & 35-44 & Female & White & >10 years & Bachelor's & \$100 - 150k & Part-Time \\ \hline
Petsitter-5 & 25-34 & Male & White & 0.5-1 year & Master's & \$100 - 150k & Part-Time \\ \hline
\end{tabular}
{\small
\caption{Participant Demographics: Workers engaged with Uber (drivers), Rover (petsitters) and Upwork (freelancers) platforms. \\
* \textit{Driver-4 dropped out after onboarding due to concerns that his participation would violate Uber policies.} \\
\textsuperscript{+} \textit{Driver-5 dropped out after onboarding due to personal reasons, preventing him from actively uploading data.}}
\label{participants}
}

\end{table*}

%% file: story_stats.tex
\begin{table*}[h]
\begin{tabular}{cc|ccc|c}
\cline{3-5}
 &  & \multicolumn{3}{c|}{\textbf{Authors's Work Contexts}} &  \\ \cline{3-6} 
 &  & \multicolumn{1}{c|}{\textit{Driver Stories}} & \multicolumn{1}{c|}{\textit{Petsitter Stories}} & \textit{Freelancer Stories} & Total \\ \cline{2-6} 
\multicolumn{1}{c|}{} & Total Authored & \multicolumn{1}{c|}{15} & \multicolumn{1}{c|}{11} & 1 & 27 \\ \cline{2-6} 
\multicolumn{1}{c|}{} & Mean Stories / User & \multicolumn{1}{c|}{2.143} & \multicolumn{1}{c|}{2.2} & 0.5 & N/A \\ \hline
\multicolumn{1}{|c|}{} & \textit{From Drivers} & \multicolumn{1}{c|}{13} & \multicolumn{1}{c|}{\textbf{10}} & 0 & 23 \\ \cline{2-6} 
\multicolumn{1}{|c|}{} & \textit{From Petsitters} & \multicolumn{1}{c|}{\textbf{10}} & \multicolumn{1}{c|}{10} & 0 & 20 \\ \cline{2-6} 
\multicolumn{1}{|c|}{\multirow{-3}{*}{\textbf{\# Likes}}} & \textit{From Freelancers} & \multicolumn{1}{c|}{\textbf{1}} & \multicolumn{1}{c|}{0} & 0 & 1 \\ \hline
\multicolumn{1}{|c|}{} & \textit{Workers Only} & \multicolumn{1}{c|}{2} & \multicolumn{1}{c|}{1} & 0 & 3 \\ \cline{2-6} 
\multicolumn{1}{|c|}{} & \textit{Policymakers Only} & \multicolumn{1}{c|}{0} & \multicolumn{1}{c|}{1} & 0 & 1 \\ \cline{2-6} 
\multicolumn{1}{|c|}{} & \textit{Workers + Policymakers} & \multicolumn{1}{c|}{1} & \multicolumn{1}{c|}{0} & 0 & 1 \\ \cline{2-6} 
\multicolumn{1}{|c|}{\multirow{-4}{*}{\textbf{Share to}}} & \textit{Workers + Policymakers + Advocates} & \multicolumn{1}{c|}{12} & \multicolumn{1}{c|}{9} & 1 & 22 \\ \hline
\multicolumn{1}{|c|}{} & \textit{Strategies} & \multicolumn{1}{c|}{10} & \multicolumn{1}{c|}{8} & 0 & 18 \\ \cline{2-6} 
\multicolumn{1}{|c|}{\multirow{-2}{*}{\textbf{Story Type}}} & \textit{Issues} & \multicolumn{1}{c|}{5} & \multicolumn{1}{c|}{3} & 1 & 9 \\ \hline
\end{tabular}
\caption{
Story statistics across platforms. 
\textit{Workers of all platforms expressed interests (via likes) for a comparable number of Stories in their domains as in others --- e.g., drivers liked 10 stories from petsitters, in addition to 13 stories from other drivers}}
\label{story_stats}
\end{table*}

%% file: summary_stats.tex
\begin{table*}[h]
\begin{tabular}{ccccccc}
\textbf{} &
  \textbf{\begin{tabular}[c]{@{}c@{}}\# Shared \\ Stories\end{tabular}} &
  \textbf{\begin{tabular}[c]{@{}c@{}}\# Total Words \\ in Stories\end{tabular}} &
  \textbf{\begin{tabular}[c]{@{}c@{}}\# Liked \\ Stories\end{tabular}} &
  \textbf{\begin{tabular}[c]{@{}c@{}}\# Income \\ Uploads\end{tabular}} &
  \textbf{\begin{tabular}[c]{@{}c@{}}\# Expense \\ Uploads\end{tabular}} &
  \textbf{\begin{tabular}[c]{@{}c@{}}\# Trends \\ Visits\end{tabular}} \\ \cline{2-7} 
\multicolumn{1}{c|}{\textit{Average}} & 1.93 & 231  & 3  & 8.57 & 1.42 & 4.5 \\
\multicolumn{1}{c|}{\textit{Median}}  & 1    & 108  & 3  & 6.5  & 1    & 4.5 \\
\multicolumn{1}{c|}{\textit{Max}}              & 5    & 1493 & 11 & 41   & 7    & 9   \\
\multicolumn{1}{c|}{Total}            & 27   & 3235 & 42 & 120  & 20   & 63 
\end{tabular}
\caption{Descriptive Statistics on Stories, Uploads and Trends}
\label{summary_stats}
\end{table*}

%% file: tags.tex
\begin{table*}[h]
\begin{tabular}{c|ccc|c|cc|c|ccc}
\multicolumn{1}{l|}{} & \multicolumn{3}{c|}{\textbf{Usage in Authored Stories}} & \multicolumn{1}{l|}{\multirow{2}{*}{\begin{tabular}[c]{@{}l@{}}Total \\ Usage\end{tabular}}} & \multicolumn{2}{c|}{\textbf{Story Type}} & \multicolumn{1}{l|}{\multirow{2}{*}{\begin{tabular}[c]{@{}l@{}}Total \\ Liked\end{tabular}}} & \multicolumn{3}{c|}{\textbf{Liked Stories}} \\ \cline{2-4} \cline{6-7} \cline{9-11} 
\multicolumn{1}{l|}{} & \textit{Driver} & \textit{Petsitter} & \textit{Freelancer} & \multicolumn{1}{l|}{} & \textit{Strategy} & \textit{Issue} & \multicolumn{1}{l|}{} & \textit{Driver} & \textit{Petsitter} & \multicolumn{1}{c|}{\textit{Freelancer}} \\ \hline
\textit{\textbf{safety}} & \multicolumn{1}{c|}{5} & \multicolumn{1}{c|}{5} & 0 & 10 & \multicolumn{1}{c|}{7} & 3 & 19 & \multicolumn{1}{c|}{11} & \multicolumn{1}{c|}{8} & \multicolumn{1}{c|}{0} \\ \hline
\textit{\textbf{fair pay}} & \multicolumn{1}{c|}{5} & \multicolumn{1}{c|}{1} & 0 & 6 & \multicolumn{1}{c|}{4} & 2 & 5 & \multicolumn{1}{c|}{3} & \multicolumn{1}{c|}{2} & \multicolumn{1}{c|}{0} \\ \hline
\textit{\textbf{care-giving}} & \multicolumn{1}{c|}{1} & \multicolumn{1}{c|}{4} & 0 & 5 & \multicolumn{1}{c|}{3} & 2 & 9 & \multicolumn{1}{c|}{5} & \multicolumn{1}{c|}{4} & \multicolumn{1}{c|}{0} \\ \hline
\textit{\textbf{stress}} & \multicolumn{1}{c|}{1} & \multicolumn{1}{c|}{3} & 0 & 4 & \multicolumn{1}{c|}{2} & 2 & 6 & \multicolumn{1}{c|}{4} & \multicolumn{1}{c|}{2} & \multicolumn{1}{c|}{0} \\ \hline
\textit{\textbf{technology}} & \multicolumn{1}{c|}{3} & \multicolumn{1}{c|}{1} & 0 & 4 & \multicolumn{1}{c|}{4} & 0 & 10 & \multicolumn{1}{c|}{5} & \multicolumn{1}{c|}{4} & \multicolumn{1}{c|}{1} \\ \hline
\textit{\textbf{other}} & \multicolumn{1}{c|}{1} & \multicolumn{1}{c|}{2} & 1 & 3 & \multicolumn{1}{c|}{2} & 2 & 3 & \multicolumn{1}{c|}{1} & \multicolumn{1}{c|}{2} & \multicolumn{1}{c|}{0} \\ \hline
\textit{\textbf{ratings}} & \multicolumn{1}{c|}{1} & \multicolumn{1}{c|}{2} & 0 & 3 & \multicolumn{1}{c|}{1} & 2 & 6 & \multicolumn{1}{c|}{3} & \multicolumn{1}{c|}{3} & \multicolumn{1}{c|}{0} \\ \hline
\textit{\textbf{work time}} & \multicolumn{1}{c|}{0} & \multicolumn{1}{c|}{2} & 0 & 2 & \multicolumn{1}{c|}{2} & 0 & 3 & \multicolumn{1}{c|}{1} & \multicolumn{1}{c|}{2} & \multicolumn{1}{c|}{0} \\ \hline
\textit{\textbf{algorithms}} & \multicolumn{1}{c|}{0} & \multicolumn{1}{c|}{1} & 0 & 1 & \multicolumn{1}{c|}{0} & 1 & 2 & \multicolumn{1}{c|}{1} & \multicolumn{1}{c|}{1} & \multicolumn{1}{c|}{0} \\ \hline
\textit{\textbf{discrimination}} & \multicolumn{1}{c|}{1} & \multicolumn{1}{c|}{0} & 0 & 1 & \multicolumn{1}{c|}{1} & 0 & 1 & \multicolumn{1}{c|}{1} & \multicolumn{1}{c|}{0} & \multicolumn{1}{c|}{0} \\ \hline
\textbf{Total} & \multicolumn{1}{c|}{18} & \multicolumn{1}{c|}{21} & 1 & \textbf{39} & \multicolumn{1}{c|}{26} & 14 & \textbf{64} & \multicolumn{1}{c|}{35} & \multicolumn{1}{c|}{28} & 1
\end{tabular}
\caption{Tag Statistics Across Platforms}
\label{tags}
\end{table*}

%% file: appendixA.tex
\appendix
\section{Appendix A: Mock ups, Protocols \& Features }
\begin{figure*}[h!]  
\subsubsection{Mock-up examples}\label{A.1.1}
The screenshots below display examples of mock-up screens used to gather iterative feedback during pilot testing: Home (a), Story Sharing (b), Work Planner (c) and Collective Insights (d)
    \centering
    \includegraphics[width=\linewidth]{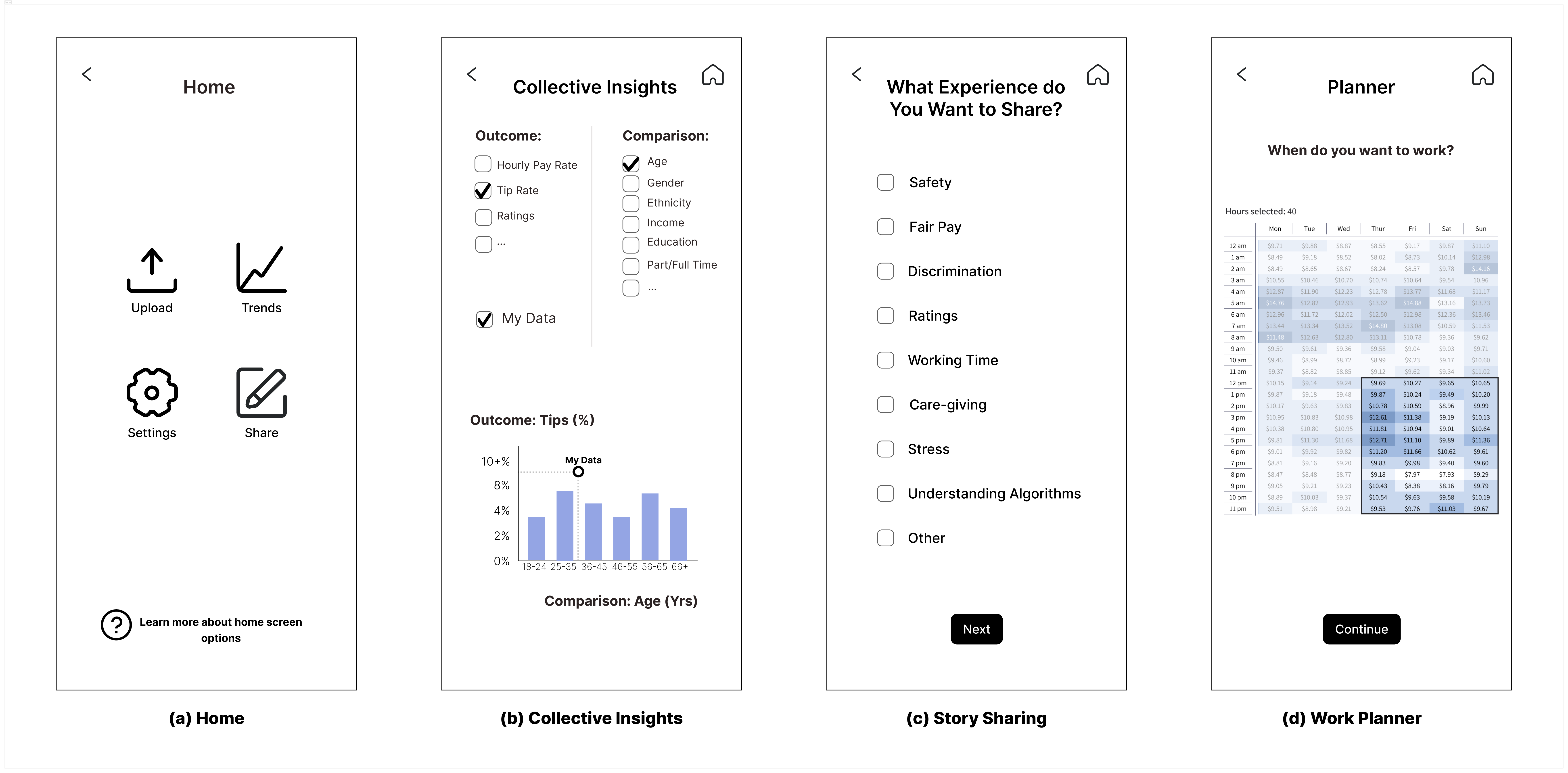}
    \label{mock-ups}
\end{figure*}
\subsection{Iterative Pilot Testing}
As overviewed in Section \ref{iterative}, participant feedback from the think-aloud sessions informed iterative refinements of the Gig2Gether prototype. These sessions were designed to evaluate the usability and functionality of essential features, including registration, data uploading, work planning, data review, qualitative input, and data-sharing preferences. Participants were prompted to articulate their thoughts throughout the tasks, enabling the research team to uncover usability issues, user preferences, and areas for improvement. Below is the list of questions and Figma mock-ups used during these sessions.
\subsubsection{Think-aloud questions}\label{thinkaloud}
\begin{itemize}
    \item Do you prefer signing up with a phone number or an email? Why?
    \item How comfortable were you with providing demographic information?
    \item Did any questions cause discomfort?
    \item Is there additional information you would need to feel comfortable completing the process?
    \item Do you think the registration process was too long?

    \item How easy or difficult was it to upload your data?
    \item Which version (mobile or web) did you prefer for this task? Why?
    \item Do you currently track data for your gigs? How do you typically do it?
    \item How does this process fit into your existing data collection habits?
    \item Would you use this app to contribute data in the future? Why or why not?

    \item How easy or difficult was it to use the 'trend' function?
    \item Which version (mobile or web) do you prefer for analyzing data? Why?
    \item Did you find the 'my trend' feature useful?
    \item What additional insights or trends would be helpful for your work?

    \item How easy or difficult was it to plan your work?
    \item Was the summary of earnings and hours useful?
    \item Did the planner feel intuitive to use? Why or why not?
    \item What additional features would help you better plan your work?

    \item How easy or difficult was it to use the 'share' function?
    \item What did you think of the available tags? Are there other tags you would like to see?
    \item Would you use this feature to share your story? Why or why not?
    \item What would you find most helpful to see in other workers' stories?

    \item How easy was it to set your sharing preferences?
    \item What data expiration settings do you prefer?
    \item Do you have any suggestions for improving location granularity settings?
    \item What other controls would you like over your data?
\end{itemize}

\subsection{Field Evaluation}
\subsubsection{Exit Interview questions}\label{exits}
\begin{itemize}
    \item How useful do you find the system? 
    \item Can you describe specific features or aspects that you find particularly helpful or unhelpful?
    \item How likely would you recommend this tool to other friends who are gig workers, if you have any who are workers?
    \item How likely are you to use this app to contribute data in the future? What factors would influence your decision?
    \item Given your experiences contributing so far, how comfortable would you share this information with policymakers and advocates?
    \item Can you reflect on your experiences of uploading your data? What parts of the process were straightforward, and where did you encounter difficulties?
    \item What times of the day did you usually upload? How did this fit into your work schedule?
    \item How does the feed compare to other forums? Do you feel more or less willing to share?
\end{itemize}

\subsection{Details of Gig2Gether Features}
\label{details}
\paragraph{Share Story} \label{share_story}
Each story must 1) be shared as a strategy or issue, 2) be associated with at least one tag, 3) contain story content via a title or textual description, and 4) have a selection of desired viewing audience -- this can include other worker users of the system, policymakers, advocates, or be entirely private (i.e. visible only to themselves). Optionally, workers can include an image or video to provide additional context. See the share story page on the left of Figure \ref{overview}(a).

\paragraph{Story Feed} The story feed provides a place for workers to exchange stories with peers on Gig2Gether. 
At registration time, users are advised to choose a username that will be viewable to other users of the system, and each post is associated with the user only through the username. Posts can be filtered by the story type (Issue or Strategy), as well as by work platform (currently Uber, Rover or Upwork). Gig2Gether allows for cross-{platform} user interaction -- users can currently view and ``like'' posts via thumbs-up buttons. Commenting is currently unsupported, in the absence of an established moderation structure.
The feed is chronologically ordered -- most recent posts appear first; an example can be found via the right side of Figure \ref{overview}(a).

\paragraph{Income} 
For Rover and Upwork users, Gig2Gether currently only supports manual data entry. 
In the income form, a worker can upload information pertaining to time spent, earnings (including the platform cut and tips), as well as information specific to job types, such as time spent travelling to house sits (Rover) and experience levels (Upwork). 
An example for the Rover manual upload is shown at \ref{overview}(d).

Uber users can manually upload data about Trips or upload CSVs that contain platform-collected data about their trips.
The \textit{Trip entry form} gathers information on the time spent, income, distance travelled, Uber fees, surge fees, as well as other specific items detailed in a Trip receipt. 
Finally, Uber allows drivers to download CSV files containing information on lifetime trips, payments and app analytics. Workers have a space to keep track of such information with Gig2Gether, offering a more expedited way of seeing personal work trends.

\paragraph{Expenses} Workers can manually input details about
incurred expenses related to gigs. To add an entry, users must enter the date and cost amount, while fields such as expense type, description and a photo uploads are optional for their own notetaking. Refer to \ref{overview}(d) for the expense upload page for Rover workers.

\paragraph{Personal Trends} 
To stay informed about earning patterns and work hours, workers can overview earnings, expense, hourly earning rates and hours worked in the ``My Trends'' page. Based on income and expense entries that users uploaded (process described in Section \ref{upload}), workers can view hourly and weekly earning trends, daily earnings by month, as well as summary statistics such as hourly pay and worked hours. The design of the hourly and calendar data visualizations in ``My Trends'' were informed by the personal data probes (in particular the Hourly and Calendar probes) from \citet{zhang2023stakeholder}. The Personal Trends page is displayed in \ref{overview}(b).

\paragraph{Collective Trends} In addition to personal metrics, workers can also view aggregate information about other Gig2Gether users via the ``Collective Insights'' page. At the time the study was conducted, this page is populated only with mock data rather than real data that workers inputted to protect the privacy of our small pool of test users. However, the page does include charts and options for dimensions of input (hourly income rate, tipping rate, and ratings) as well as demographic information to breakdown each dimension by (age, gender, ethnicity, income, education, tenure, and part/full-time work). Users can additionally compare their own data point against any breakdowns displayed. Refer to \ref{overview}(e) to view the Collective Insights page.

%% file: appendixB.tex
\section{Appendix B: Scams} \label{scams}
Scams were, unfortunately, shared experiences that resonated with workers of all platforms. Although ``true'' scams occur rarely on Rover, Petsitter-4 described how they usually take the form of a ``classic check scheme'' where the scammer claims ``they're going to send you a check for \$500 and tell you to buy something and send back whatever is extra''. Manipulations of hours or number of pets involved are more commonplace, where clients would change hours to ``get charged less for a boarding or a house set, and they can manipulate the number of animals \dots the cost comes out to us [as sitters]'' (Petsitter-4). On Uber, Driver-7 described getting phone calls from fake numbers claiming to be Uber support who tries to assign him `` `a ride to a very important person. So we need to confirm your identity' ''. The scammer would then proceed to ask for their phone number to which send a 4-digit code, which they'll then use to access the drivers' account. Meanwhile, Freelancer-1 related how she enjoyed reading about others' \textit{Stories} of ``scams \dots cause there's quite a few of them on Upwork''.

%% file: appendixC.tex
\section{Appendix C: Stories Workers Shared}
\begin{figure}[H]
    \includegraphics[width=\linewidth]{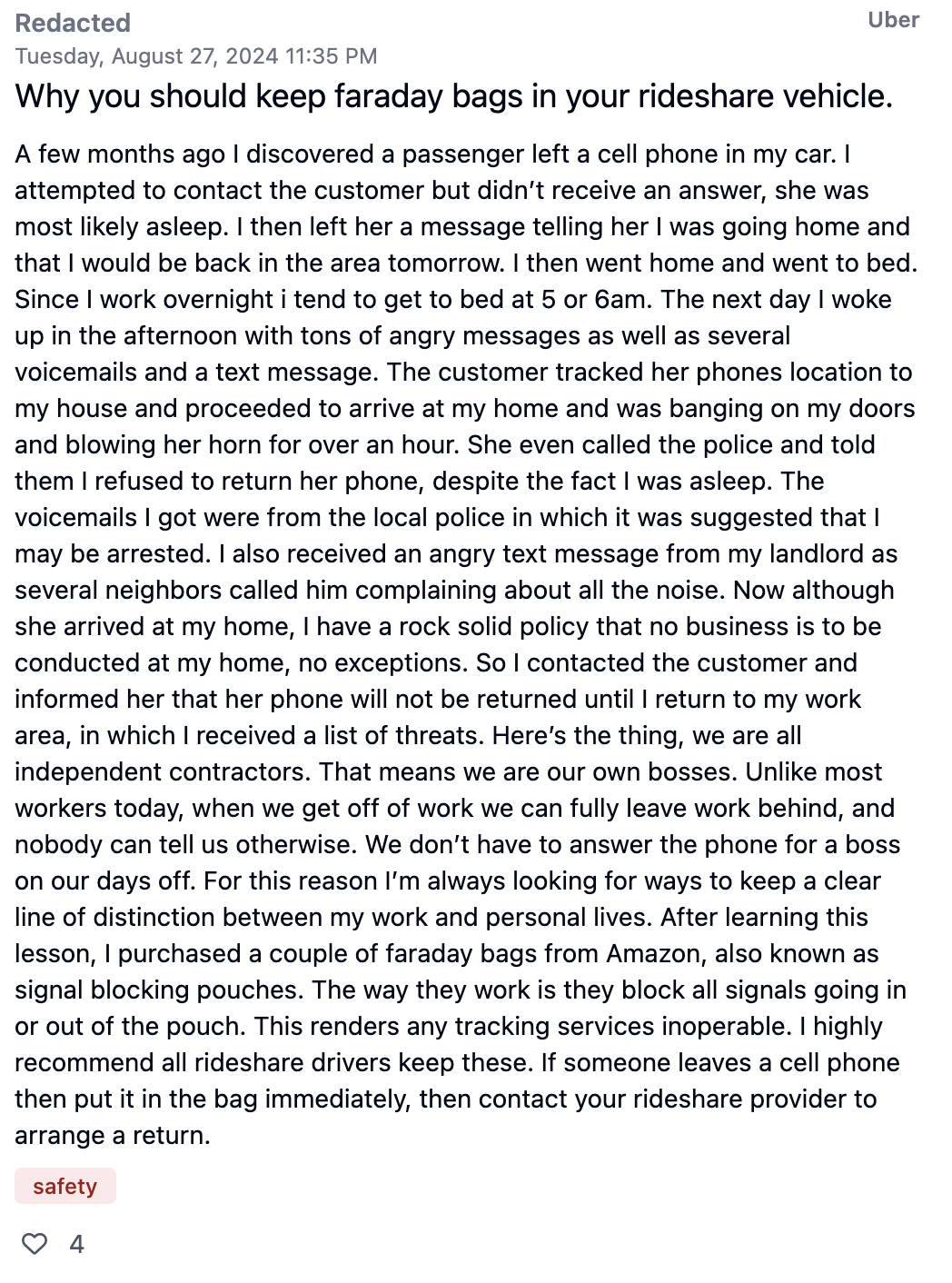}
    \caption{Driver-1's Strategy: Self-Protecting from a Trespassing Passenger}
    \label{phone_customer}
\end{figure}
\begin{figure}[H]
    \includegraphics[width=\linewidth]{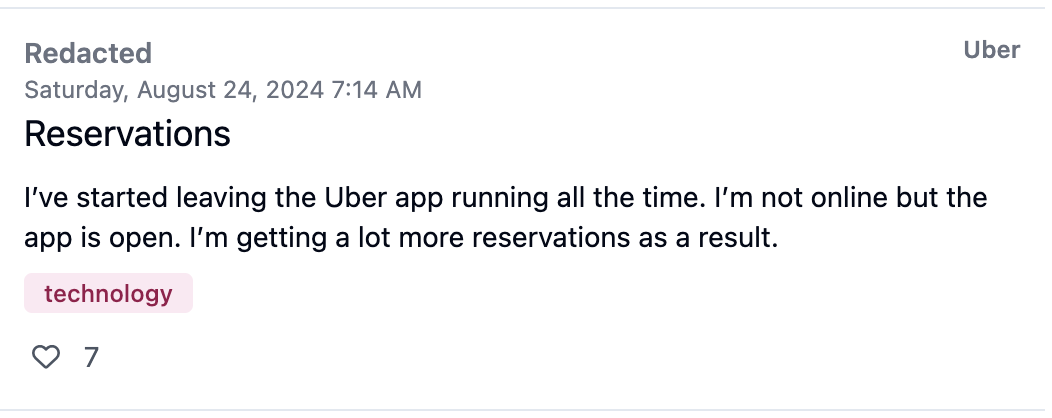}
    \caption{Driver-2's Strategy: More Reservation Assignments}
    \label{reservation}
\end{figure}
\begin{figure}[H]
    \includegraphics[width=\linewidth]{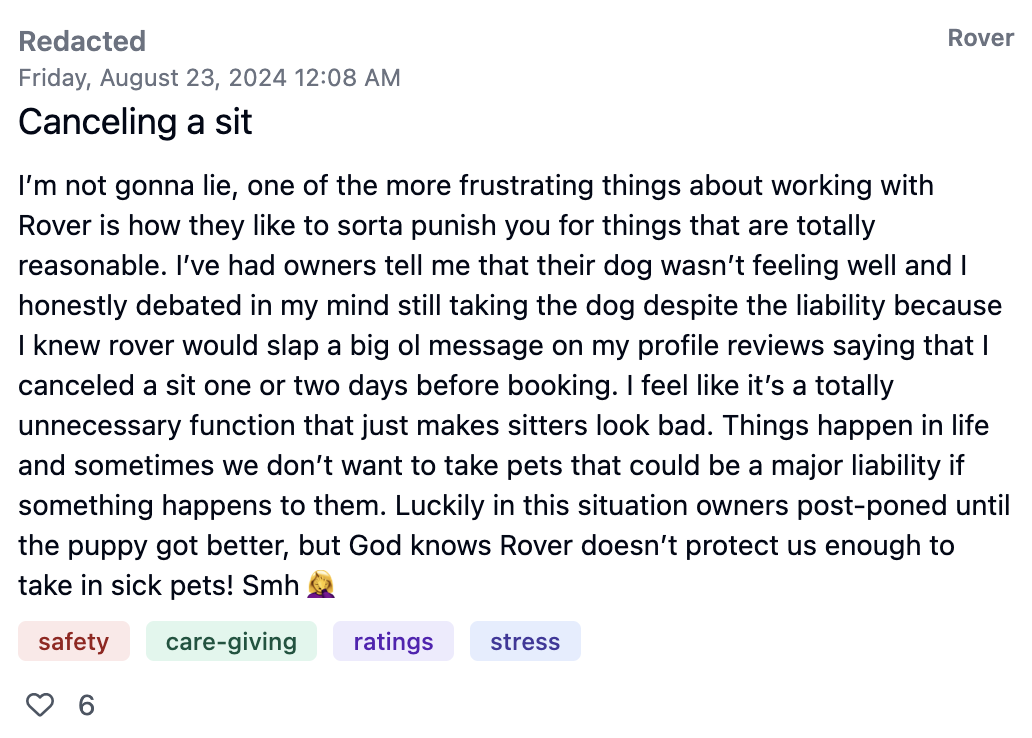}
    \caption{Petsitter-2's Fear of Cancellations}
    \label{power}
\end{figure}
\begin{figure}[H]
    \includegraphics[width=\linewidth]{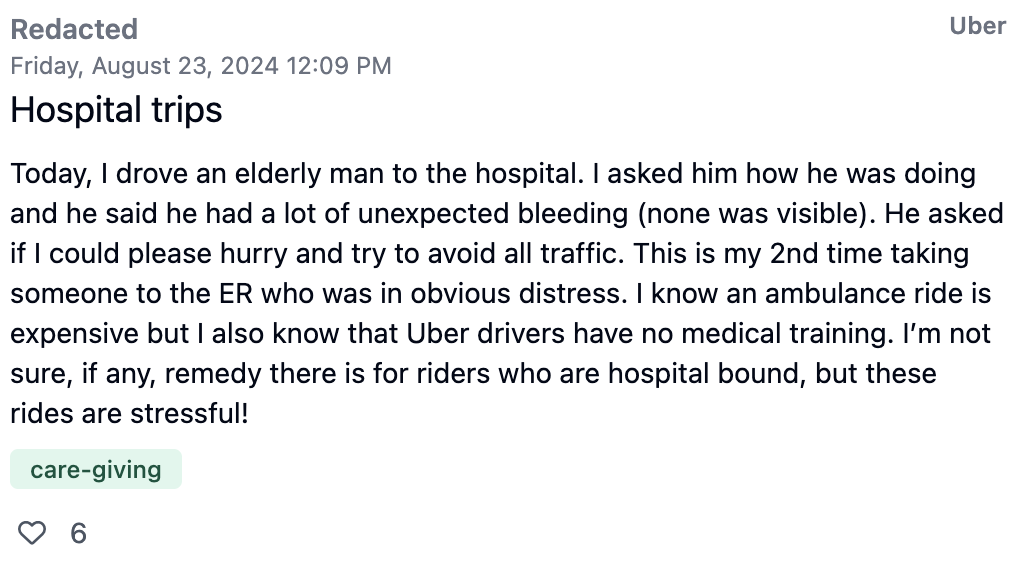}
    \caption{Driver-2's Stressful Ride to the ER}
    \label{d2_hospital}    
\end{figure}
\begin{figure}[H]
    \includegraphics[width=\linewidth]{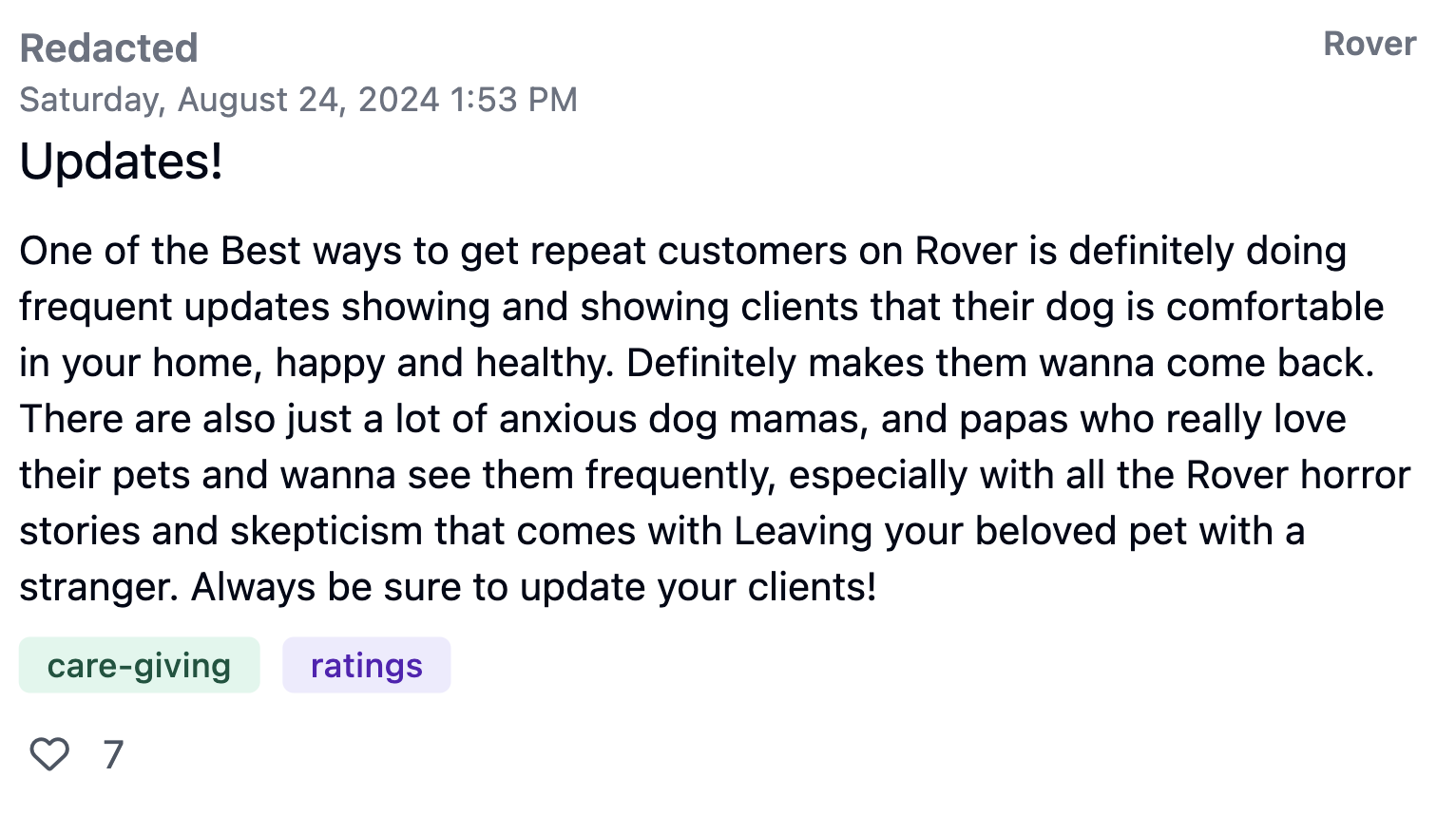}
    \caption{Petsitter-2's Strategy: Repeat Customers}
    \label{p2_repeat}
\end{figure}
\begin{figure}[H]
    \includegraphics[width=\linewidth]{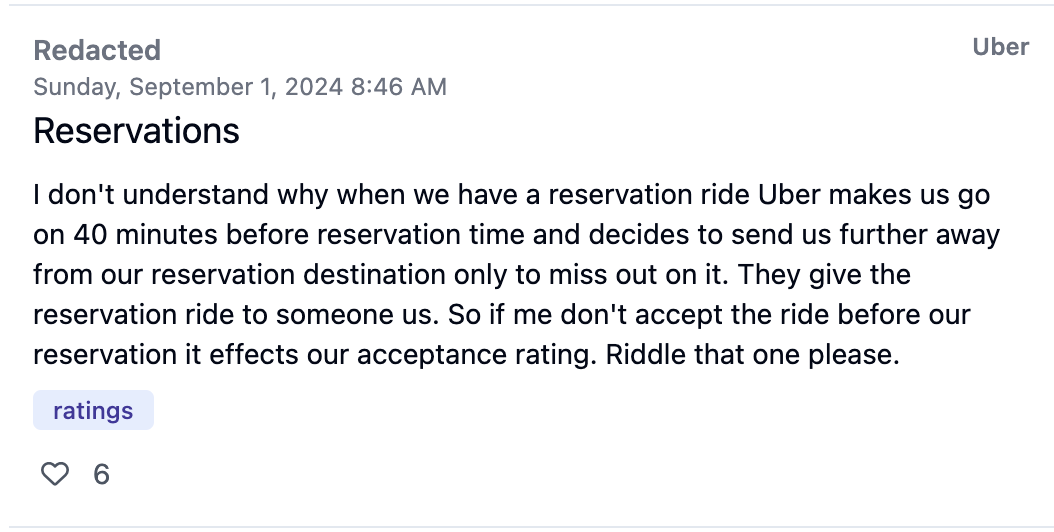}
    \caption{Driver-3 Frustration on Reservation Assignments}
    \label{d3_power}
\end{figure}
\begin{figure}[H]
    \includegraphics[width=\linewidth]{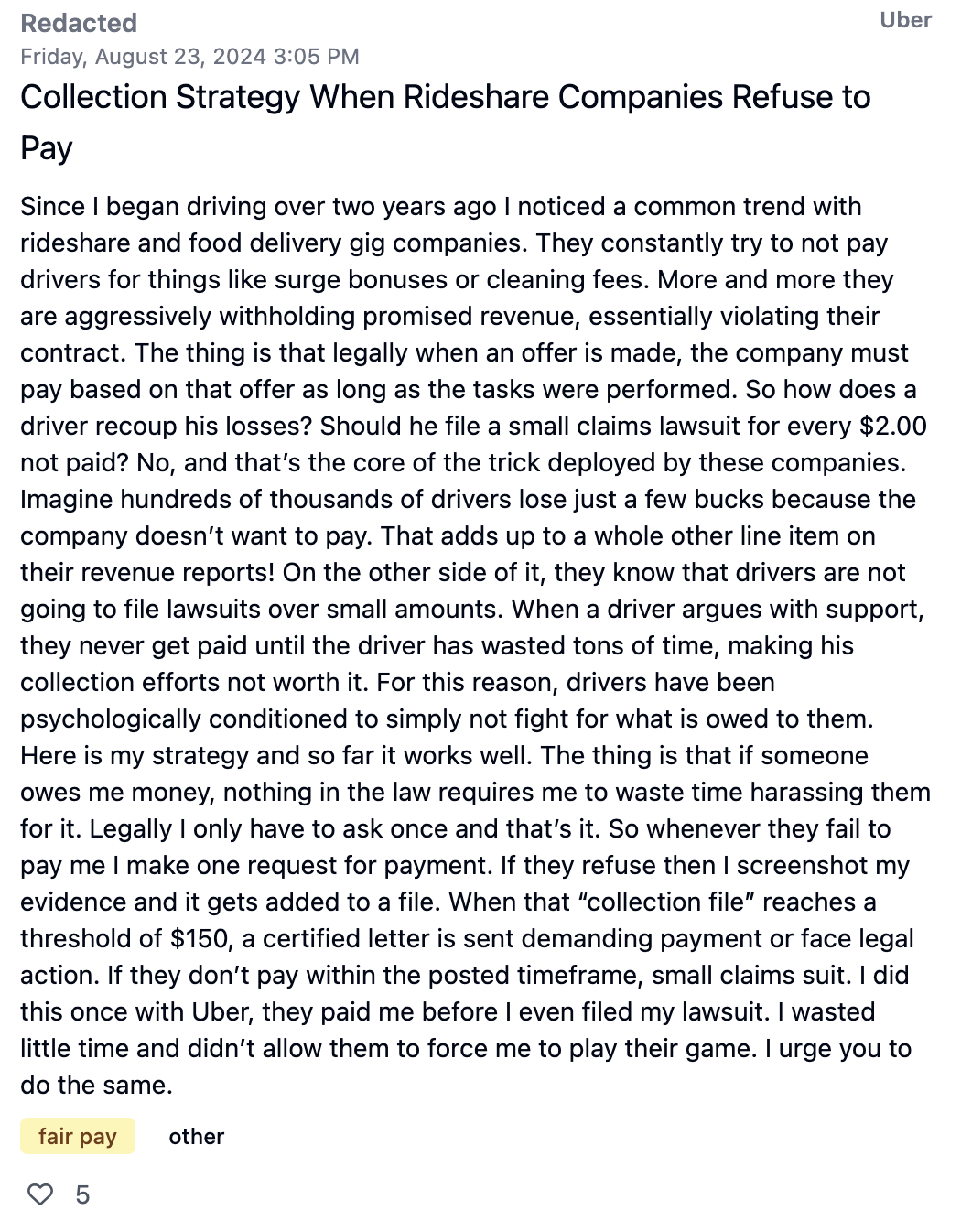}
    \caption{Driver-1's strategy: Small Claims Lawsuits}
    \label{small-claims}    
\end{figure}
\begin{figure}[H]
    \includegraphics[width=\linewidth]{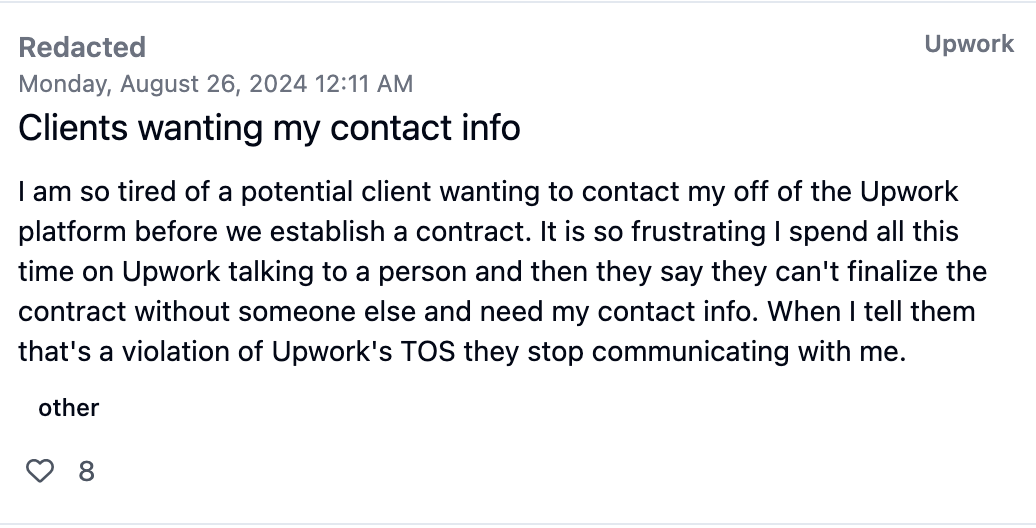}
    \caption{Freelancer-1's Issue: Client-worker Power Dynamics}
    \label{f1_power}
\end{figure}
\begin{figure}[H]
        \includegraphics[width=\linewidth]{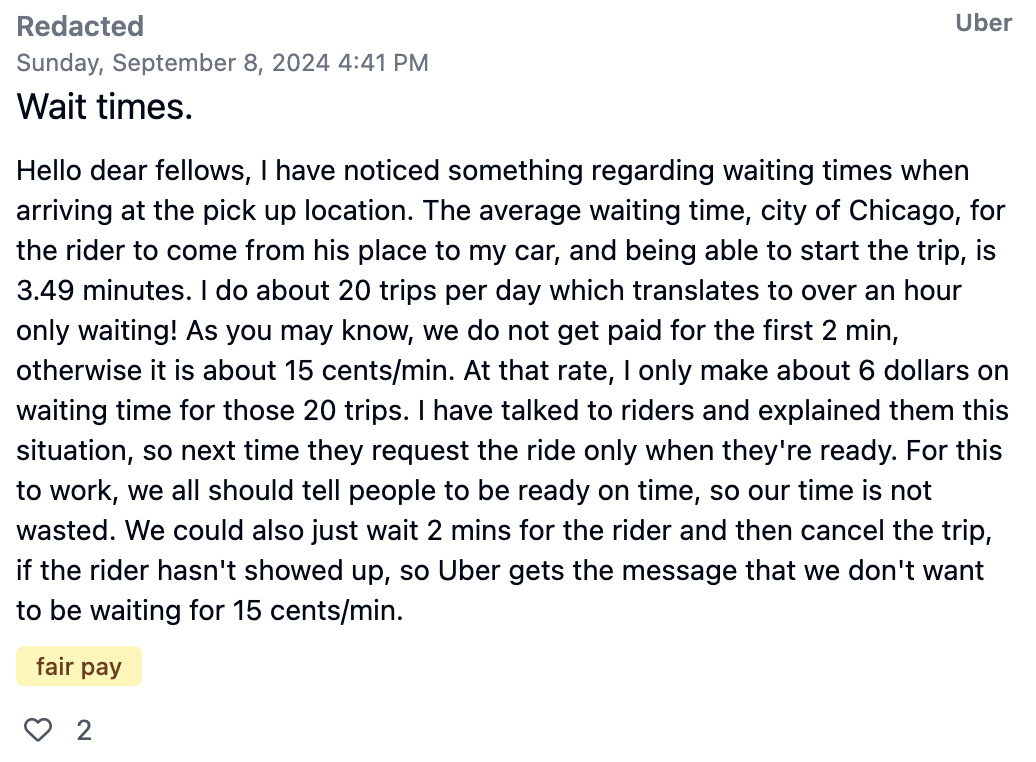}
    \caption{Driver-7's Observations on Unpaid Time }
    \label{d7_platform}
\end{figure}
\begin{figure}[H]
        \includegraphics[width=\linewidth]{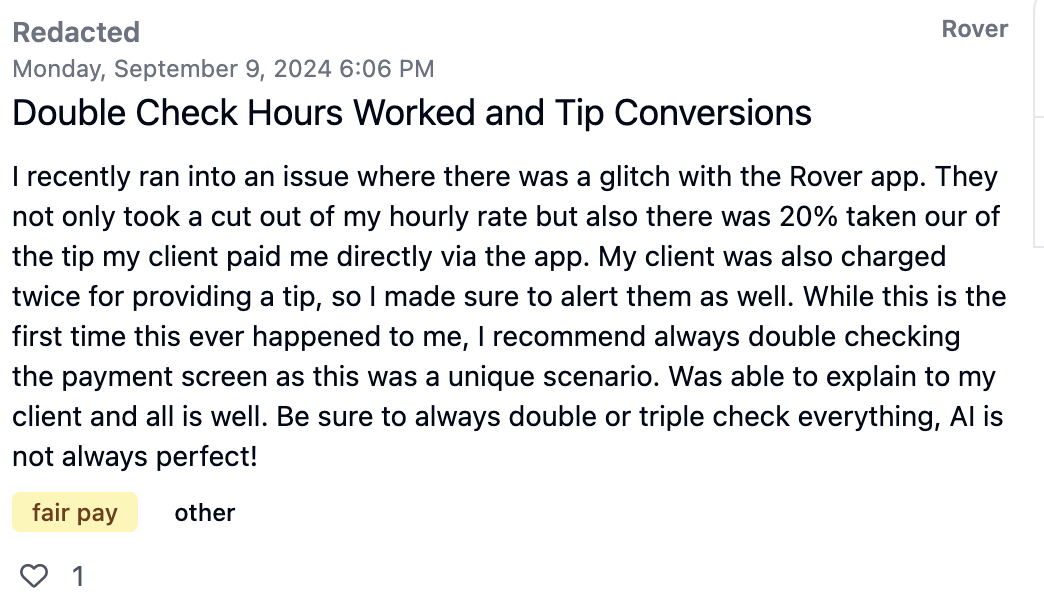}
    \caption{Petsitter-5's Issue: Double Charged Tips}
    \label{p5_platform}
\end{figure}
\begin{figure}[H]
        \includegraphics[width=\linewidth]{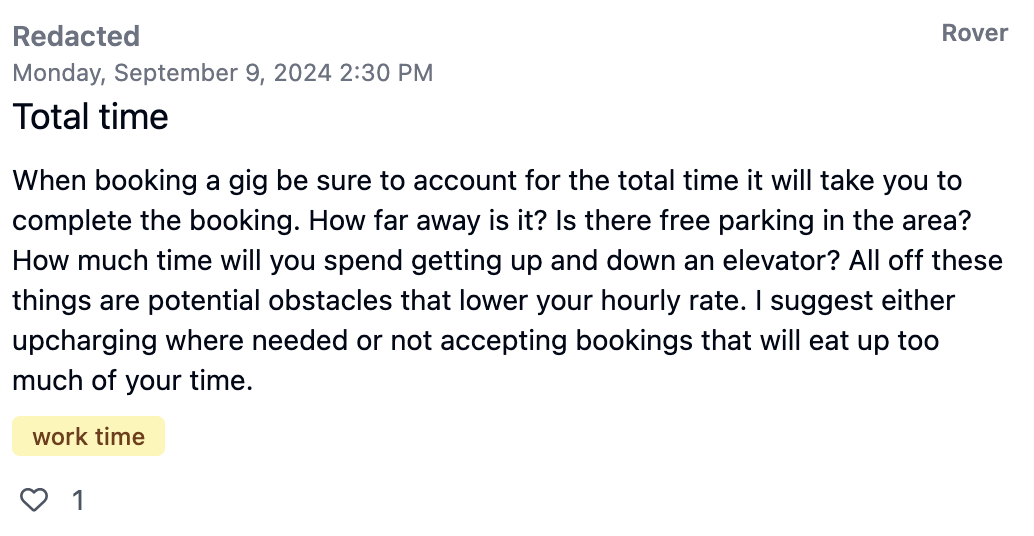}
    \caption{Petsitter-4's Strategy to Record Unpaid Tasks}
    \label{p4_unpaid}
\end{figure}
\begin{figure}[H]
    \includegraphics[width=\linewidth]{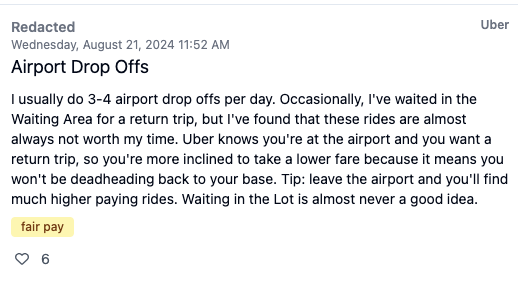}
    \caption{Driver-2's Strategy: Airport Rides}
    \label{d2_airport}
\end{figure}